\documentclass[aps,reprint,prl,nofootinbib,twocolumn]{revtex4}
\usepackage{amsmath,amssymb}
\usepackage{graphicx,subfigure}
\usepackage{color,multirow}
\usepackage{booktabs}
\usepackage[colorlinks,linkcolor=magenta,anchorcolor=cyan,citecolor=blue,plainpages=false]{hyperref}

\hypersetup{colorlinks=true,
    breaklinks=true,
    pdfstartview=Fit,
    linkcolor=blue,
    citecolor=blue,
    urlcolor=blue}

\def\be{\begin{equation}}
	\def\ee{\end{equation}}
\def\ba{\begin{eqnarray}}
    \def\ea{\end{eqnarray}}

\begin{document}

\title{Beyond the Peak: Constraining early dark energy with the damping tail}

\author{Hao Wang$^{1} $\footnote{\href{wanghao187@mails.ucas.ac.cn}{wanghao187@mails.ucas.ac.cn}}}
\author{Gen Ye$^{1} $\footnote{\href{yegen@ucas.ac.cn}{yegen@ucas.ac.cn}}}
\author{Yun-Song Piao$^{1,2,3,4} $ \footnote{\href{yspiao@ucas.ac.cn}{yspiao@ucas.ac.cn}}}

    \affiliation{$^1$ School of Physics Sciences, University of
        Chinese Academy of Sciences, Beijing 100049, China}
        
    \affiliation{$^2$ School of Fundamental Physics and Mathematical
        Sciences, Hangzhou Institute for Advanced Study, UCAS, Hangzhou
        310024, China}

    \affiliation{$^3$ International Center for Theoretical Physics
        Asia-Pacific, Beijing/Hangzhou, China}

    \affiliation{$^4$ Institute of Theoretical Physics, Chinese
        Academy of Sciences, P.O. Box 2735, Beijing 100190, China}

    %   \date{}
    \begin{abstract}

Early dark energy is commonly characterized by the amplitude and location of its primary energy-injection peak. We show that its post-peak evolution, which we call the \emph{EDE tail}, provides an additional observable probe of the underlying scalar dynamics and can materially affect its ability to alleviate the Hubble tension. We construct a simple phenomenological framework in which distinct regions of a multiscale scalar potential govern the dynamics at different field scales. As a representative realization, we consider a potential that combines axion-like and $\alpha$-attractor-like regimes, thereby allowing the EDE tail to be reshaped through reflection and asymmetric oscillations. Confronting this framework with cosmic microwave background, baryon acoustic oscillation, and supernova observations, we find an improved fit to the high-$\ell$ CMB data together with a higher inferred bestfit $H_0$. Our results show that the EDE tail retains observable information about the underlying scalar potential that is not captured by the conventional peak parameters alone. This provides a complementary route for constructing and observationally distinguishing EDE scenarios, and potentially for further alleviating the Hubble tension.

    \end{abstract}

    \maketitle
    %\tableofcontents
    %\newpage

\textbf{\textit{Introduction.}} ---
The discrepancy between local measurements of the Hubble constant and the
value inferred from the cosmic microwave background (CMB) within the standard
$\Lambda$CDM model remains one of the most persistent challenges in cosmology
\cite{Verde:2019ivm,DiValentino:2020zio,DiValentino:2021izs,Perivolaropoulos:2021jda,Shah:2021onj,Abdalla:2022yfr,Hu:2023jqc,Verde:2023lmm,CosmoVerseNetwork:2025alb,H0DN:2025lyy,SPT-3G:2025bzu,Cuceu:2019for,Schoneberg:2019wmt,DESI:2025zgx}. If it cannot be attributed to unresolved
systematic effects\cite{Efstathiou:2020wxn,Mortsell:2021tcx,Kenworthy:2022jdh,Camarena:2022iae,Wojtak:2022bct,Riess:2023bfx,Giani:2023aor,Wojtak:2024mgg,Freedman:2024eph,Perivolaropoulos:2024yxv,Riess:2025chq,Hogas:2026urs}, this discrepancy may provide evidence for physics beyond
$\Lambda$CDM\cite{Mortsell:2018mfj,Vagnozzi:2018jhn,Yang:2018euj,Guo:2018ans,Kreisch:2019yzn,Vagnozzi:2019ezj,Visinelli:2019qqu,DiValentino:2019ffd,DiValentino:2019jae,Krishnan:2020obg,Alestas:2020mvb,Jedamzik:2020krr,Sekiguchi:2020teg,RoyChoudhury:2020dmd,Brinckmann:2020bcn,Gao:2021xnk,Marra:2021fvf,Dainotti:2021pqg,Krishnan:2021dyb,Cyr-Racine:2021oal,Anchordoqui:2021gji,Akarsu:2021fol,Ren:2022aeo,Nojiri:2022ski,Schoneberg:2022grr,Banerjee:2022ynv,deSa:2022hsh,Akarsu:2022typ,Lee:2022gzh,Khodadi:2023ezj,Bernui:2023byc,Ben-Dayan:2023rgt,Gomez-Valent:2023hov,Ruchika:2023ugh,Adil:2023exv,Frion:2023xwq,Gomez-Valent:2023uof,Akarsu:2024qiq,Giare:2024ytc,Lynch:2024gmp,Giare:2024smz,Lynch:2024hzh,Toda:2024ncp,Nozari:2024wir,Escamilla:2024xmz,RoyChoudhury:2024wri,Mirpoorian:2024fka,Li:2025owk,Lee:2025yah,Teixeira:2025czm,Wang:2025dzn,Zhang:2025dwu,Kumar:2025obb,Hogas:2025mii,Pantos:2026rpe,Bernal:2016gxb,Lemos:2018smw,Aylor:2018drw,Knox:2019rjx,Arendse:2019hev,Cai:2021weh,Keeley:2022ojz,Jiang:2024xnu,Zhou:2025kws,Pedrotti:2025ccw,Li:2026asg,Dhyani:2026trw,Zhou:2026iar,Cai:2026swf,Sabogal:2026qvy,Wang:2026jxd,Sabogal:2026ipu,Ye:2024kus,Wang:2024sgo,DeFelice:2026led,Li:2026mbh,Wang:2024hwd,Pedrotti:2026dwj,Giare:2024syw,Wang:2026kbg,Wang:2024dka}. Among the proposed explanations, early dark energy (EDE) has
emerged as a leading possibility by introducing a transient energy component
before recombination and thereby allowing a larger CMB-inferred value of
$H_0$ \cite{Karwal:2016vyq,Poulin:2018cxd,Smith:2019ihp,Kamionkowski:2022pkx,Poulin:2023lkg,Agrawal:2019lmo,Lin:2019qug,Niedermann:2019olb,Berghaus:2019cls,Sakstein:2019fmf,Ye:2020btb,Zumalacarregui:2020cjh,Ballesteros:2020sik,Braglia:2020iik,Ballardini:2020iws,Gogoi:2020qif,Braglia:2020bym,Gonzalez:2020fdy,CarrilloGonzalez:2020oac,Braglia:2020auw,Adi:2020qqf,Karwal:2021vpk,Jiang:2021bab,Gomez-Valent:2021cbe,Poulin:2021bjr,Niedermann:2021vgd,Wang:2022jpo,Smith:2022hwi,Gomez-Valent:2022bku,Escudero:2022rbq,Herold:2022iib,Goldstein:2023gnw,Eskilt:2023nxm,Odintsov:2023cli,Sharma:2023kzr,Fu:2023tfo,Pedreira:2023qqt,Khalife:2023qbu,Forconi:2023hsj,Wang:2024jug,Giare:2024akf,Chatrchyan:2024xjj,Jiang:2024nha,Forconi:2025cwp,Piras:2025eip,Stahl:2025czl,Jiang:2025hco,Forconi:2025zzu,Smith:2025grk,Yashiki:2025loj,Toda:2025kcq,Yin:2026gss,Gonzalez-Fuentes:2026rgu,Jhaveri:2026bla,Du:2026qtq,Bella:2026zuk,Efstathiou:2023fbn,McDonough:2023qcu,Hill:2020osr,Poulin:2025nfb,SPT-3G:2025vyw,Planck:2020olo,Carloni:2026yut,Cerqueira:2026epl,Zhou:2026jmn,Sengupta:2026gyg,Giare:2026gdn,Seto:2024cgo,Adi:2025hyj,Smith:2025icl,Wang:2026aez,Wang:2025djw}. These studies also have important implications for research on the very early Universe\cite{Ye:2021nej,Ye:2022efx,Wang:2024tjd,Peng:2025tqt,Jiang:2022uyg}.

% Current observations, however, have increasingly challenged the ability of
% EDE to fully resolve the Hubble tension. In particular, analyses of the
% Planck PR4/NPIPE temperature and polarization data with the CamSpec and
% HiLLiPoP likelihoods place tighter bounds on the EDE fraction than those
% obtained with the conventional Planck PR3 Plik likelihood
% \cite{McDonough:2023qcu,Efstathiou:2023fbn}. Much of this constraining power
% comes from the high-$\ell$ CMB spectra, where EDE leaves characteristic
% imprints on the acoustic peaks and damping tail. The resulting bounds limit
% the upward shift in $H_0$ allowed within the conventional axion-like
% scenario, with CMB lensing and large-scale-structure observations imposing
% further restrictions \cite{Hill:2020osr}. Although analyses including ACT
% and SPT data have reported a preference for a nonzero EDE component and a
% larger value of $H_0$ \cite{Smith:2022hwi,Jiang:2022uyg}, the viability of
% EDE remains sensitive to the adopted data combination and, especially, to
% the small-scale CMB measurements.

The EDE scenario features a short period of non-negligible energy fraction near matter-radiation equality which quickly becomes negligible again afterwards due to, for example, faster redshift rate than radiation and matter. The study of EDE has primarily focused on the amplitude and location of the main energy-injection peak. Although these quantities characterize the leading reduction of the sound horizon, they do not uniquely determine the subsequent evolution.

\begin{figure}
	\includegraphics[width=1\columnwidth]{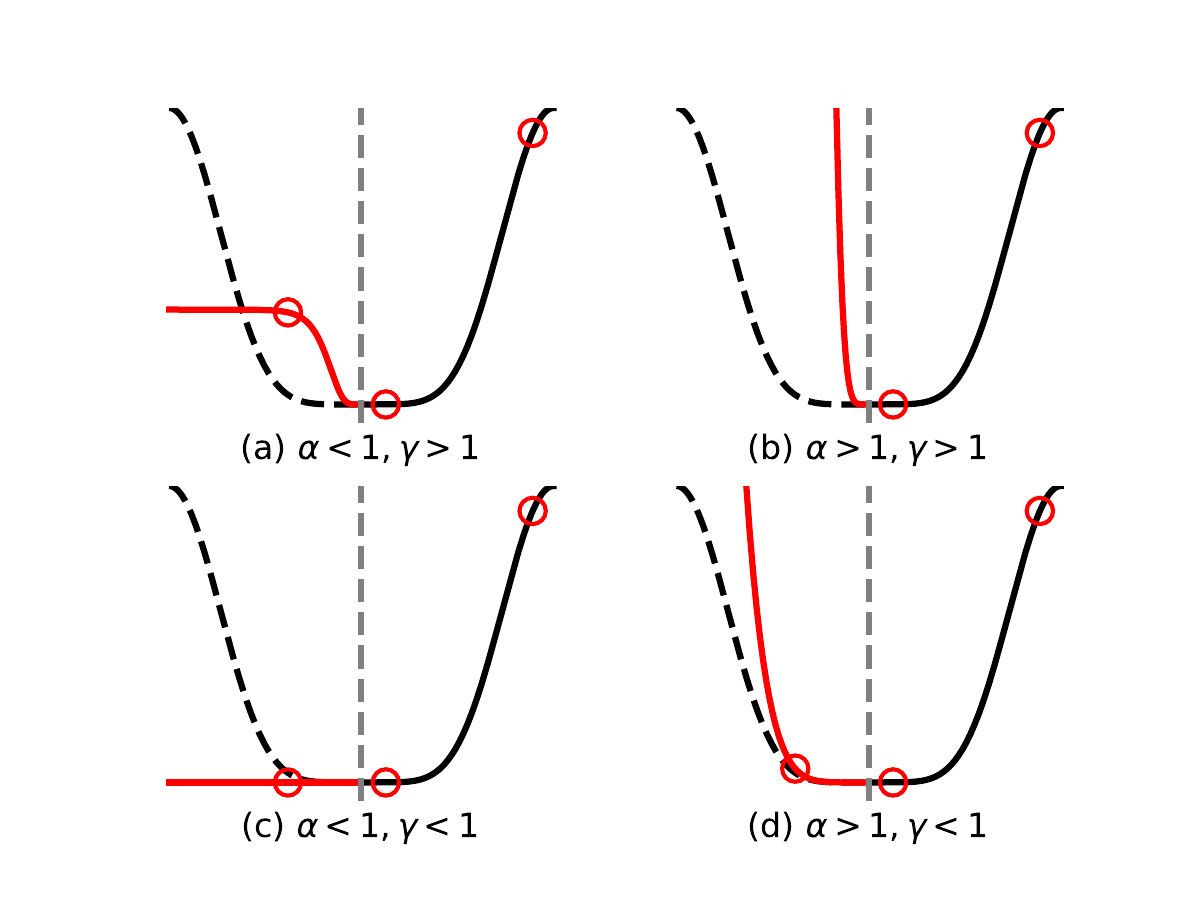}
	\caption{Illustrative potentials for AxionEDE (black dashed) and AtR (red solid) in different regions of the $(\alpha,\gamma)$ parameter space. Each panel shows a representative potential shape associated with the indicated region; the correspondence between the potential morphology and the parameter ranges is qualitative rather than exact. The four panels illustrate distinct dynamical behaviors: (a) transient slow-roll, (b) reflection, (c) runaway and (d) asymptotic $\phi^6$ oscillations.}
	%\caption{Illustrative potentials for axionEDE, AtN, and AtR. The AtN and AtR examples are chosen to have the same $\gamma$, giving the same leading behavior near the minimum, $V(\phi)\propto\gamma\phi^6$, as well as the same maximum potential height.}
	\label{vphi}
\end{figure}

It is known that the acoustic peaks and damping tail of the high-$\ell$ CMB spectra are modified by the EDE. It has been shown that improved high-$\ell$ CMB temperature and polarization data from Planck PR4 places tighter upper bounds on the EDE fraction and $H_0$ than those obtained with the conventional Planck PR3 \cite{McDonough:2023qcu,Efstathiou:2023fbn}, with CMB lensing and large-scale-structure observations imposing further restrictions \cite{Hill:2020osr}. On the other hand, alternative high-$\ell$ temperature and polarization data from the ground-based CMB measurements SPT and ACT substantially extends the measured high $\ell$ multiples. Analyses including ACT and SPT data have reported a preference for a nonzero EDE component and a larger value of $H_0$ \cite{Smith:2022hwi,Jiang:2022uyg}. Besides shift in the cosmological parameters, the main impact of EDE on the high-$\ell$ CMB spectra is through its sub-horizon perturbation dynamics, corresponding to the field evolution after the initial energy-injection peak, which we will refer to as the \emph{EDE tail} in the following.

In this paper, to investigate different EDE tail structures against observations, we adopt a phenomenological parametrized potential that can realize different tail structures with different parameters. We consider a potential which is axion-like \cite{Poulin:2018cxd,Smith:2019ihp} at $\phi>0$ and $\alpha$-attractor-like \cite{Carrasco:2015rva,Akrami:2017cir,Ye:2020oix} at $\phi<0$, with the potential minimum located at $\phi=0$. The axion-like regime ensures the initial field evolution until the main EDE peak behaves as standard AxionEDE \cite{Poulin:2018cxd}, while the $\alpha$-attractor-like regime, depending on the model parameters, supports qualitatively distinct EDE tails, including reflection, transient slow-roll, runaway, and asymptotic $\phi^6$ oscillations, see Fig.~\ref{vphi} for an illustrative sketch. This construction allows us to study the EDE tail effects in a controlled manner. Confronting this framework with CMB, baryon acoustic oscillation, and Type Ia supernova data including the local distance-ladder calibration, we find that the data prefers the reflection branch with $\Delta\chi^2=-3.34$, mainly coming from the improved fit to the high-$\ell$ CMB spectra. A transient slow-roll branch capable of producing a secondary energy injection also remains allowed with $\Delta\chi^2=-2.95$. These results establish the EDE tail as an additional observable probe of early-universe scalar dynamics and provide a complementary direction for studying EDE as a possible resolution to the Hubble tension.

\textbf{\textit{Analysis setup.}} --- We construct a representative multiscale potential describing a transition from \textbf{A}xion \textbf{t}o $\alpha$-attracto\textbf{R} (AtR) branch:
\be
V(\phi)=
\begin{cases}
	m_a^2f_a^2[1-\cos(\phi/f_a)]^n,&\phi\geq0,\\
	(\gamma\alpha^6/2^m)m_a^2f_a^2\tanh^{2m}(\phi/\alpha f_a),&\phi<0,
\end{cases}
\ee
as illustrated in Fig.~\ref{vphi}. The field is initially frozen at $\phi_i>0$ on the axion-like branch. It then rolls toward the minimum and crosses into the $\alpha$-attractor branch, whose structure reshapes the EDE tail following the primary energy injection. Depending on the parameters of the negative-field branch, the subsequent evolution can exhibit distinct behaviors, including reflection and transient slow-roll, before the field returns to the minimum and enters an oscillatory dilution phase.

On the axion-like side, the potential is characterized by the axion mass $m_a$ and decay constant $f_a$, as in the standard AxionEDE model. We fix $n=m=3$, for which both branches behave as $\phi^6$ sufficiently close to the minimum. In particular,
\be
	V(\phi)\simeq
	\begin{cases}
		(m_a^2/8f_a^4)\phi^6,&\phi\rightarrow0^+,\\
		\gamma(m_a^2/8f_a^4)\phi^6,&\phi\rightarrow0^-.
	\end{cases}
	\label{potential_expansion}
\ee
thus all potential quantities, i.e.$V$, $V_\phi$, $V_{\phi\phi}$, relevant for the background and linear perturbation equations are continuous at $\phi=0$. Once the oscillation is confined to the local $\phi^6$ region, the cycle-averaged equation of state approaches $\langle w_\phi\rangle=1/2$, independently of the asymmetry parameter $\gamma$.

The potential gives rise to several qualitatively distinct dynamical regimes. For $\alpha\gg1$, the field typically remains within the local $\phi^6$ region, and its post-peak evolution is controlled predominantly by $\gamma$. When $\gamma\ll1$, the negative branch becomes sufficiently shallow for the field to travel far from the minimum, approaching a runaway-like limit. By contrast, for $\gamma\gg1$, the field encounters an increasingly steep wall and undergoes a sharp reflection. At the background level, such a reflection approximately reverses the field velocity and can reproduce a background evolution similar to standard AxionEDE, although the perturbation need not be identical.

For $\alpha<1$, the negative branch approaches its plateau over a relatively narrow field range. For suitable values of $\gamma$, the field can climb onto the plateau and enter a transient slow-roll phase, producing a secondary energy injection before returning to the minimum and resuming its oscillatory dilution. In some cases, the field instead remains on the plateau, causing its energy density to become excessively large at late times; such solutions are strongly disfavored by late-time observations. Otherwise, the negative branch continuously reshapes the EDE tail without producing a clearly separated secondary injection.

% As a structurally minimal comparison, we also consider an \textbf{A}xion-\textbf{t}o-axio\textbf{N} (AtN) transition model,
% \be
% V(\phi)=
% \begin{cases}
% 	m_a^2f_a^2[1-\cos(\phi/f_a)]^n,&\phi\geq0,\\
% 	\gamma\alpha^6m_a^2f_a^2[1-\cos(\phi/\alpha f_a)]^n,&\phi<0
% \end{cases}
% \ee
% where $n=3$. Both sides retain the same axion-like functional form and differ only in their characteristic field and energy scales. This potential can also support multiple dynamical behaviors while the potential departs $\sim\phi^2$ from its maximum rather than approaching an extended plateau \gyc{I do not understand this sentence}. This model therefore provides a useful reference for determining whether the observational preference for a particular post-peak dynamical branch originates from the specific structure of the potential rather than merely from the introduction of two additional degrees of freedom. \gyc{I do not understand the purpose of this AtN model. It seem to me it can also realize the 4 scenarios in Fig.1. It is not well explained why AtR is favored over AtN and most of the discussion focuses on AtR. Can we remove AtN altogether?}%An illustration for this model is also displayed in Fig.\ref{vphi}.

%prolonged slow roll requires a near-critical trajectory and therefore occupies a much narrower region of parameter space.

In standard AxionEDE, the phenomenological parameters, ${f_{\rm EDE},z_c}$, height and redshift of the main (highest) EDE energy fraction peak, are mapped to the potential parameters ${m_a,f_a}$ through a shooting mechanism. In the present model, however, the various dynamical branches make the conventional shooting procedure potentially non-unique. We therefore sample the potential parameters $m_a, f_a, \alpha, \gamma$ directly and treat $f_{\rm EDE}$ and $z_c$ as derived quantities.

Our parameter space is \{$\omega_b$, $\omega_{cdm}$, $H_0$, $\ln(10^{10}A_s)$, $n_s$, $\tau_{\rm reio}$, $\log_{10}m_a$, $\log_{10}f_a$, $\Theta_i$, $\log_{10}\alpha$, $\log_{10}\gamma$\}, where $\Theta_i\equiv\phi_{\rm ini}/f_a$ represents the initial position of the field on the potential. We assume that the field start from the axion-like branch, i.e. $\phi_{\rm ini}>0$. We adopt uniform priors for the scalar-field parameters $\log_{10}m_a\in[4,8]$, $\log_{10}f_a\in[-3,0]$, $\log_{10}\alpha\in[-2,1]$, $\log_{10}\gamma\in[-3,12]$ and $\Theta_i\in[0,3.1]$. We implement this model in a modified version of the publicly available \texttt{AxiCLASS} code\footnote{\url{https://github.com/PoulinV/AxiCLASS}} as an extension of the \texttt{CLASS} Boltzmann solver\footnote{\url{https://github.com/lesgourg/class_public}}\cite{Lesgourgues:2011re,Blas:2011rf}. Both the homogeneous scalar field and its linear perturbations are evolved explicitly, without employing an effective-fluid approximation. To accurately resolve the steep-reflection regime, we additionally impose a local integration-step condition based on the effective curvature scale $\omega_\mathrm{eff}^2(\phi)=|\partial_\phi^2V(\phi)|$. This modification improves numerical stability without changing the underlying background or perturbation equations. 

We use the Planck PR4 \texttt{CamSpec} high-$\ell$ TTTEEE\cite{Rosenberg:2022sdy} and the Planck 2018 \texttt{Commander} and \texttt{SimAll} low-$\ell$ TT and EE likelihoods, and the PR3 lensing data\cite{Planck:2018lbu,Planck:2019nip}. We also include the DESI DR2 BAO measurements\cite{DESI:2025zgx} and the Cepheid-calibrated PantheonPlusSH0ES SNe Ia sample\cite{Scolnic:2021amr,Riess:2021jrx,Brout:2022vxf}, whose calibration corresponds approximately to a Gaussian constraint $H_0=73.04\pm1.04$km/s/Mpc. We perform the cosmological parameter estimation using the Monte Carlo Markov chain (MCMC) method implemented in the public package \texttt{cobaya}\cite{Torrado:2020dgo} and adopt a Gelman-Rubin convergence criterion with a threshold $R-1<0.05$\cite{Gelman:1992zz}. We perform the profile-likelihood analysis with the publicly available \texttt{Prospect} package\cite{Holm:2023uwa}, which is particularly well suited for exploring and minimizing over our complex, multimodal parameter space.

\textbf{\textit{Results.}} ---
Our main constraints are summarized in Tab.~\ref{post}, with the key posterior distributions shown in Fig.~\ref{mc}. We find that the mean value of $H_0$ in the AtR model is shifted upward by about $\simeq0.2$km/s/Mpc, accompanied by a more concentrated posterior peak. Although the shift in $H_0$ is relatively modest, the axion parameters $m_a$ and $f_a$ undergo much more pronounced changes once the shape of the potential for $\phi<0$ is modified. This suggests a nontrivial redistribution of the likelihood within the posterior region relative to AxionEDE, rather than simply a uniform displacement toward larger $H_0$. 

The posterior simultaneously allows regions with $\alpha<1$ and $\alpha>1$, corresponding to qualitatively different dynamical branches, including slow-roll-like evolution and sharp reflection. For sufficiently large $\alpha$, after initial thawing the field evolution is confined to the $\sim\phi^6$ region around the potential minima, whose shape is controlled predominantly by $\gamma$. Therefore, in this case, $\alpha$ itself becomes only weakly constrained. 
% By contrast, in the analogous AtN model the $\alpha<1$ region is disfavored with a larger $1\sigma$ lower bound. This comparison suggests that the preference for $\alpha<1$ is associated with the specific structure of the potential, rather than being a generic consequence of introducing additional potential parameters.

\begin{table*}[htbp]
	\centering
	\begin{tabular}{lcccc}
		\toprule
		Parameters			&$\Lambda$CDM&AxionEDE					&AtR						&AtR(a)\\
		\midrule
		$100\omega_b$		&$2.240(2.240)^{+0.012}_{-0.012}$&$2.254(2.248)^{+0.018}_{-0.021}$		&$2.260(2.244)^{+0.012}_{-0.015}$	&$2.250$\\
		$\omega_{cdm}$		&$0.116(0.117)^{+0.001}_{-0.001}$&$0.128(0.129)^{+0.003}_{-0.003}$		&$0.129(0.131)^{+0.003}_{-0.003}$	&$0.131$\\
		$H_0$				&$68.51(68.39)^{+0.28}_{-0.28}$&$71.41(71.92)^{+0.94}_{-0.77}$		&$71.63(72.19)^{+0.79}_{-0.71}$	&$72.29$\\
		$\ln(10^{10}A_s)$		&$3.049(3.049)^{+0.013}_{-0.014}$&$3.061(3.060)^{+0.013}_{-0.017}$		&$3.064(3.062)^{+0.016}_{-0.015}$	&$3.065$\\
		$n_s$				&$0.969(0.965)^{+0.004}_{-0.003}$&$0.986(0.988)^{+0.006}_{-0.006}$		&$0.988(0.989)^{+0.006}_{-0.006}$	&$0.990$\\
		$\tau_{\rm reio}$	&$0.060(0.060)^{+0.006}_{-0.007}$&$0.057(0.052)^{+0.006}_{-0.008}$		&$0.058(0.055)^{+0.007}_{-0.008}$	&$0.056$\\
		\midrule
		$\log_{10}f_a$		&--&$-0.842(-0.811)^{+0.091}_{-0.087}$	&$-0.743(-0.733)^{+0.110}_{-0.121}$	&$-0.730$\\
		$\log_{10}m_a$		&--&$5.666(5.548)^{+0.161}_{-0.225}$	&$5.557(5.665)^{+0.160}_{-0.255}$	&$5.638$\\
		$\Theta_i$			&--&$2.835(2.847)^{+0.120}_{-0.050}$	&$2.662(2.753)^{+0.300}_{-0.071}$	&$2.858$\\
		$\log_{10}\alpha$	&--&--						&$>-0.139(-0.500)$	&$-0.610$\\
		$\log_{10}\gamma$	&--&--							&$3.529(4.122)^{+2.605}_{-2.240}$	&$4.801$\\
		\midrule
		$f_\mathrm{EDE}$	&--&$0.105(0.115)^{+0.028}_{-0.021}$		&$0.109(0.129)^{+0.026}_{-0.021}$	&$0.072$\\
		$\log_{10}z_c$		&--&$3.632(3.571)^{+0.072}_{-0.117}$		&$3.587(3.650)^{+0.077}_{-0.122}$	&$3.832$\\
		\midrule
		$\Omega_m$			&$0.298(0.299)^{+0.003}_{-0.004}$&$0.297(0.295)^{+0.002}_{-0.003}$		&$0.297(0.296)^{+0.003}_{-0.004}$	&$0.296$\\
		$S_8$				&$0.801(0.803)^{+0.008}_{-0.007}$&$0.826(0.825)^{+0.009}_{-0.011}$		&$0.829(0.829)^{+0.010}_{-0.011}$	&$0.832$\\
		\bottomrule
	\end{tabular}
	\caption{\label{post} Marginalized posterior means and $1\sigma$ uncertainties for
		the $\Lambda$CDM, AxionEDE, and AtR models obtained from Planck
		NPIPE high-$\ell$ CMB, Planck 2018 low-$\ell$ and lensing,
		DESI DR2 BAO, and PantheonPlusSH0ES data.
		Values in parentheses denote the corresponding best-fit values.
		The AtR(a) column presents an illustrative double-injection
		solution rather than a marginalized constraint.}
\end{table*}

\begin{figure}
\includegraphics[width=1\columnwidth]{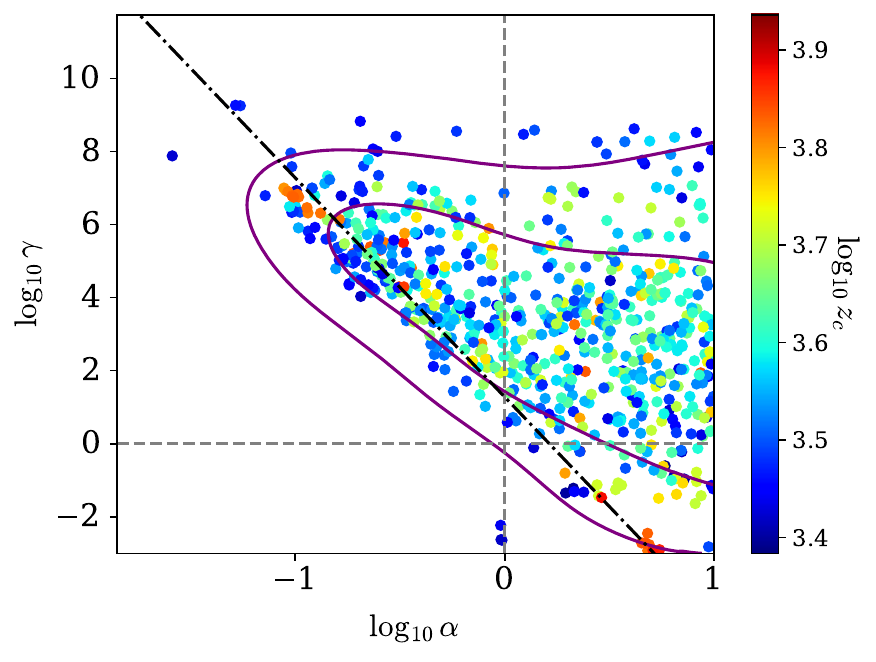}
\caption{Marginalized posterior in the $\log_{10}\alpha$--$\log_{10}\gamma$ plane for AtR, with samples colored by the derived $\log_{10}z_c$. The contours enclose the $68\%$ and $95\%$ credible regions, and the gray dashed lines mark $\alpha=1$ and $\gamma=1$. The black dot-dashed line denotes the constant-$\lambda$ relation in (\ref{deg}) for $\lambda\simeq1.6$, a value close to that at the minimum. Samples near this line separate into $\log_{10}z_c\simeq3.8$ and $3.4$, corresponding to the primary and secondary EDE injections.}
\label{alpgamma}
\end{figure}

The posterior in Fig.~\ref{alpgamma} exhibits a distinct low-$\alpha$ branch. In the $\alpha<1$ region, a clear degeneracy between $\gamma$ and $\alpha$ emerges, approximately following
\be
\log_{10}\gamma +6\log_{10}\alpha\equiv\log_{10}\lambda\sim\text{const.}\label{deg}
\ee
The $\lambda\equiv\gamma\alpha^6$ characterizes the height of the $\alpha$-attractor branch relative to the axion potential and therefore controls how easily the field can enter a slow-roll regime. The EDE tail becomes a well-resolved secondary energy injection only within a narrow region around this degeneracy line. In this case, depending on parameters and initial condition, the second energy injection peak can be higher than the first peak, resulting in a $z_c$ that is distinct from the rest of the points, which appears as orange and red points in Fig.~\ref{alpgamma}.

\begin{figure}
	\includegraphics[width=1\columnwidth]{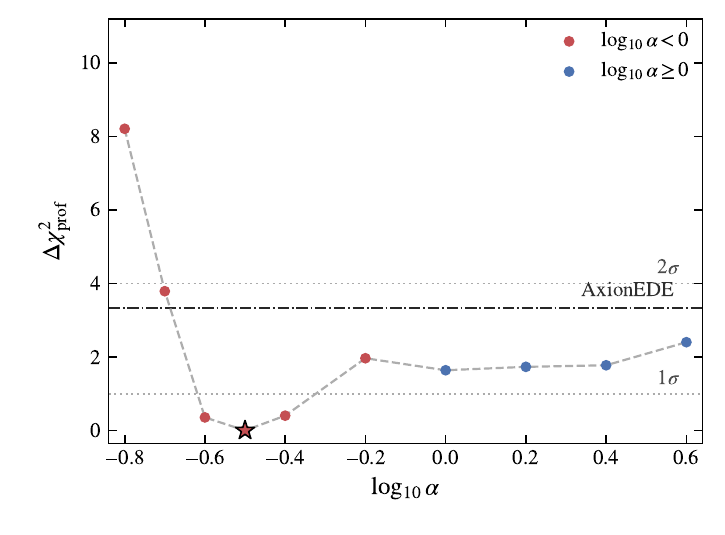}
	\caption{Profile likelihood of $\log_{10}\alpha$ for the AtR model. The red and blue points correspond to the $\log_{10}\alpha<0$ and $\log_{10}\alpha\geq0$ branches, respectively. The star marks the best evaluated AtR point. The dotted horizontal lines indicate the $1\sigma$ and $2\sigma$ thresholds, $\Delta\chi^2_{\rm prof}=1$ and $4$. The black dot-dashed line denotes the minimum $\chi^2$ obtained for AxionEDE. The profile is normalized to the best evaluated AtR point.}
	\label{profile}
\end{figure}

\begin{figure*}
	\includegraphics[width=1\columnwidth]{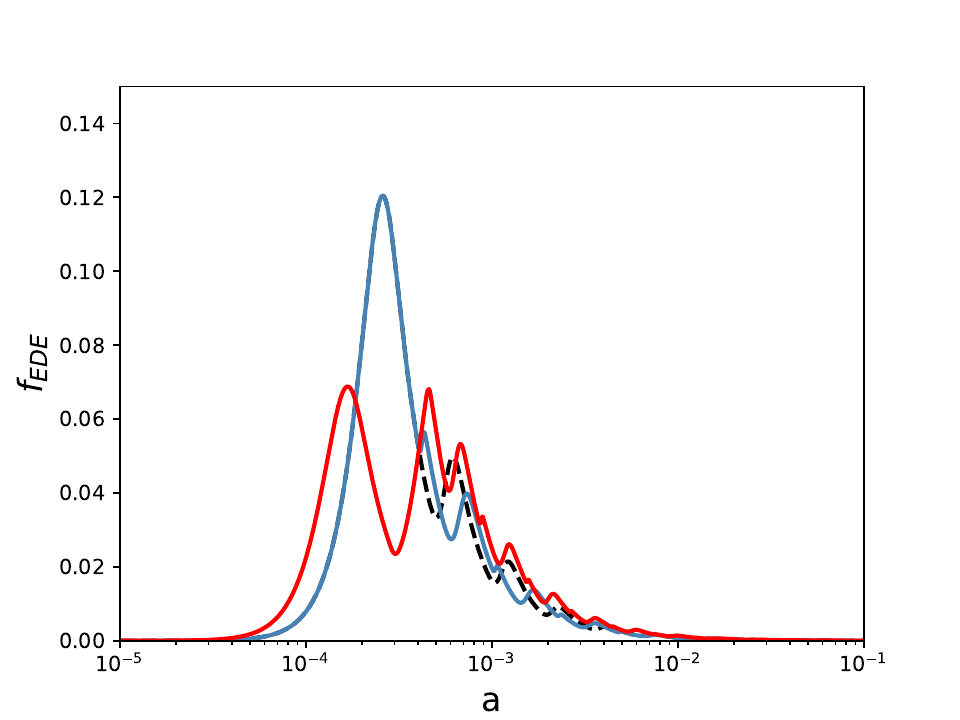}
	\includegraphics[width=1\columnwidth]{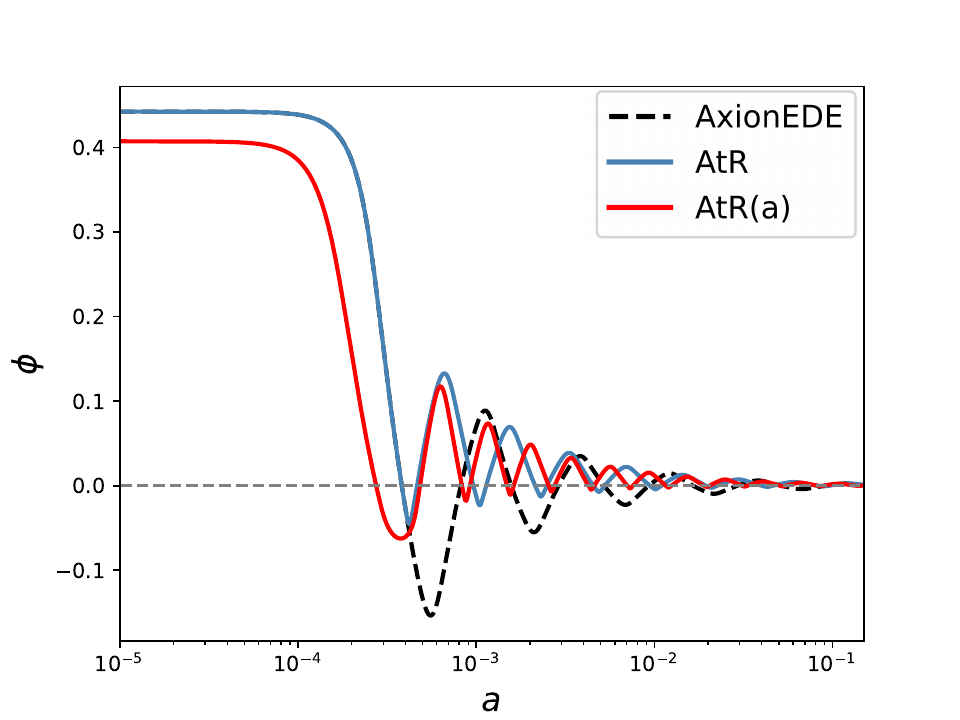}
	\caption{Background evolution for AxionEDE, the AtR minimum, and an illustrative double-injection solution associated with the potential morphology in Fig.~\ref{vphi}(a). The left panel shows the EDE energy fraction $f_{\rm EDE}$, and the right panel shows the scalar-field evolution. The AxionEDE curve is evaluated using the same cosmological parameters as the AtR minimum.}
	\label{fede}
\end{figure*}

As commented previously, the large $\alpha$ region is weakly constrained compared with the $\alpha<1$ region, raising the issue of volume effects in the posteriors. To further explore the complicated and multimodal parameter space, we perform a profile-likelihood analysis in $\log_{10}\alpha$ in Fig.~\ref{profile} by fixing the value of $\log_{10}\alpha$ at each point and minimizing the other parameters. Profile likelihood has the advantage of avoiding the volume effect arising from the weak constraint on $\alpha>1$. The profile likelihood exhibits a pronounced low-$\alpha$ minimum (global bestfit), while the high-$\alpha$ points, corresponding to branches (b) and (d) in Fig.~\ref{vphi}, form a broader, shallower valley which is repelled at $1.5\sigma$ with $\Delta\chi^2\simeq2$. This indicates that current data is able to distinguish different types of EDE tails. The bestift point corresponds to a special case of branch (a) in Fig.~\ref{vphi} where the field is reflected by the steep potential step at $\phi<0$ and does not reach the plateau. It has been pointed out in Ref.\cite{Poulin:2025nfb} that existence of a second EDE injection peak might help improve fit to CMB by studying multiple EDE fields. Here, by further examining the posterior points near the bestfit, we notice the double-peak scenario can also be realized by as single field with potential from branch (a), where the field reaches the plateau at $\phi<0$ and creates a second EDE peak, but with slightly worse $\chi^2$ than the global bestfit. For comparison, in the following we refer to the model with a second EDE energy as AtR(a). Table~\ref{chi2} lists the per-data $\chi^2$ for the global bestfit and an illustrative solution of AtR(a), while Fig.~\ref{fede} shows their background evolution. For comparison, we also show AxionEDE evaluated with the same cosmological parameters as the AtR bestfit, thereby isolating the effect of the $\alpha$-attractor branch. For the global minimum, the modified potential suppresses the amplitude of the oscillatory dilution phase and induces a sharp reflection shortly after the field crosses $\phi=0$. The AtR(a) case, on the other hand, shows very different behavior where the slow-roll in the $\alpha<1$ branch develops a clear secondary energy injection following a relatively short first peak. This behavior is associated with a smooth turnaround near the shoulder of the $\alpha$-atrractor branch, while the subsequent oscillatory dilution retains an amplitude comparable to that of AxionEDE. The double-injection solution features a lower peak amplitude but a substantially broader energy-injection period.
% Although the phenomenological realization contains two potential parameters, the preferred solutions occupy an approximately one-dimensional dynamical branch. The improvement should therefore be interpreted as sensitivity to a particular deformation of the EDE tail rather than as evidence for unrestricted freedom in the full two-dimensional parameter space.

\begin{table*}[htbp]
	\centering
	\begin{tabular}{lcccc}
		\toprule
		Dataset						&$\Lambda$CDM&AxionEDE	&AtR	&AtR(a)\\
		\midrule
		NPIPE\_highl\_CamSpec.TTTEEE&10547.13&10546.79	&10544.48	&10545.20	\\
		planck\_2018\_lowl.TT		&22.77&21.03		&20.91	&20.96		\\
		planck\_2018\_lowl.EE		&397.33&395.80		&396.23	&396.39		\\
		planck\_2018\_lensing.native&9.28&9.81		&9.70	&9.46		\\
		bao.desi\_dr2.desi\_bao\_all&10.93&10.34		&10.35	&10.43		\\
		sn.pantheonplusshoes		&1485.61&1459.95	&1458.72	&1458.33	\\
		\midrule
		$\chi^2_\mathrm{tot}$		&12473.05&12443.72	&12440.39	&12440.77	\\
		$\Delta\chi^2_\mathrm{tot}$	&+29.33&0			&-3.34	&-2.95		\\
		\bottomrule
	\end{tabular}
	\caption{\label{chi2} Best-fit $\chi^2$ contributions for $\Lambda$CDM, AxionEDE, the AtR minimum, and an illustrative AtR double-injection solution [AtR(a)] associated with Fig.~\ref{vphi}(a). All models are fitted to the same combination of Planck NPIPE high-$\ell$ CMB, Planck 2018 low-$\ell$ and lensing, DESI DR2 BAO, and PantheonPlusSH0ES.}
\end{table*}

\begin{figure*}
	\includegraphics[width=1.8\columnwidth]{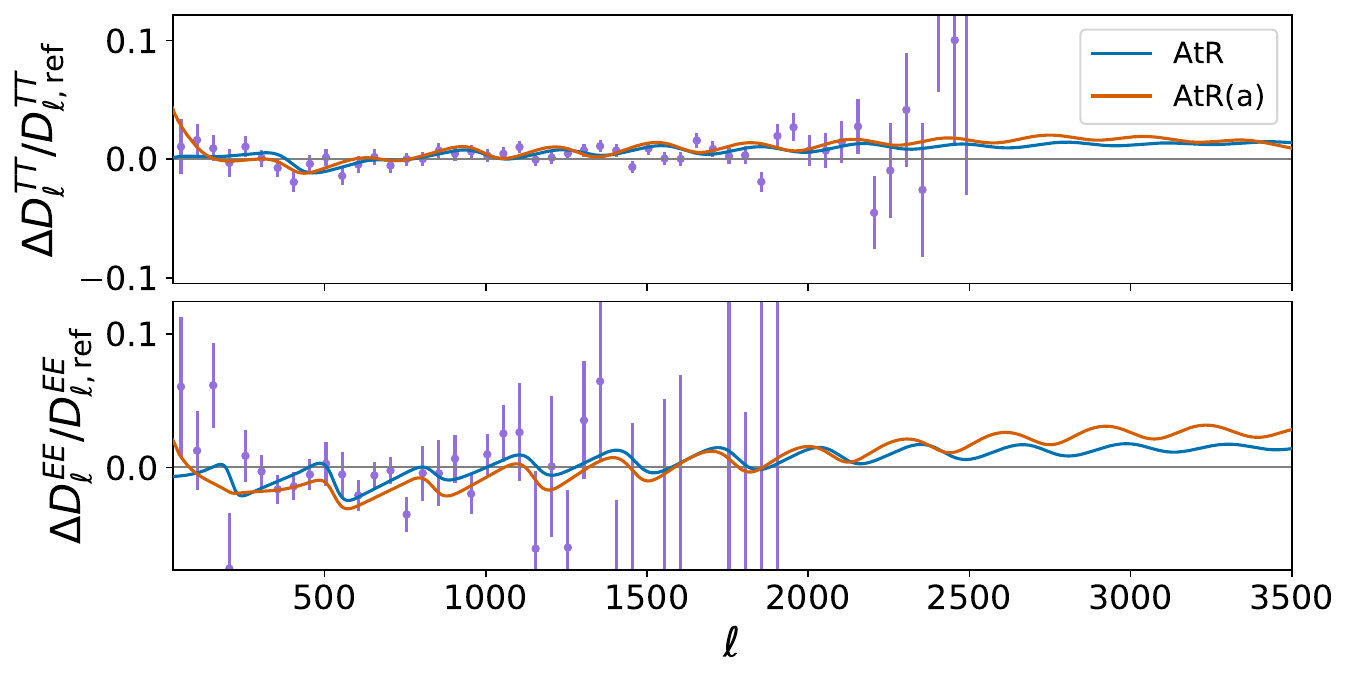}
	\caption{Relative residuals of the lensed CMB TT and EE power spectra ($30<\ell<3500$), relative to the reference AxionEDE model. Blue and vermilion curves represent AtR and AtR(a), respectively, while purple points show the binned Planck NPIPE measurements with $1\sigma$ uncertainties.}
	\label{cmb}
\end{figure*}

\begin{figure}
	\includegraphics[width=1\columnwidth]{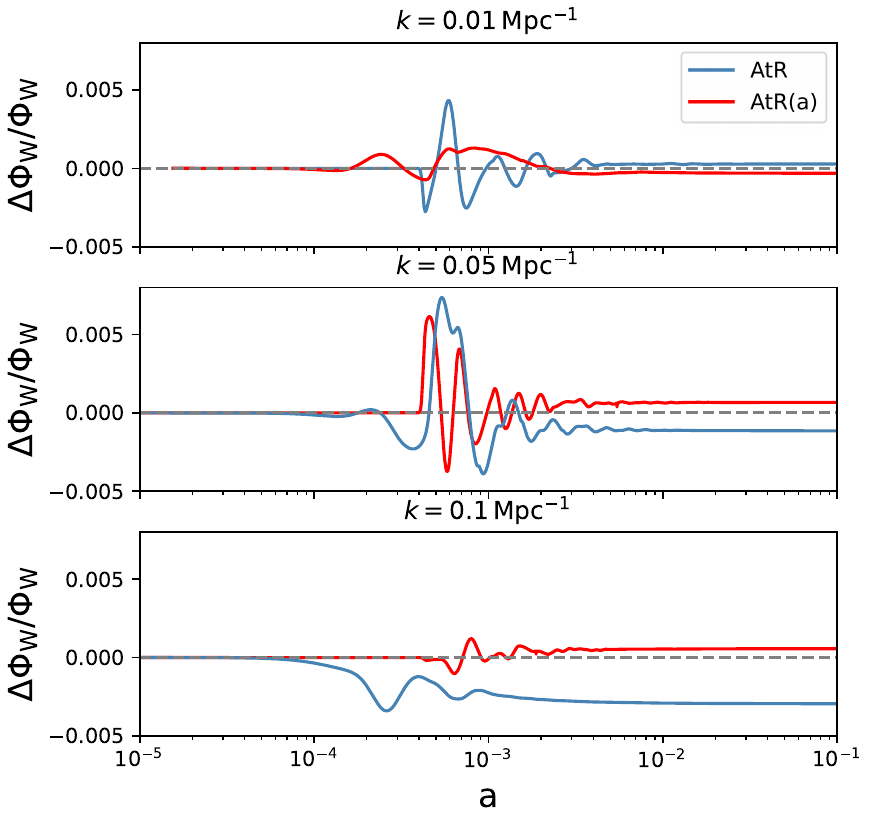}
	\caption{Relative residuals of the Weyl potential,
		$\Delta\Phi_{\rm W}/\Phi_{\rm W}$, with respect to the AxionEDE reference model, for $k=0.01$, $0.05$, and $0.1\,{\rm Mpc}^{-1}$ from top to bottom.}
	\label{weyl}
\end{figure}

Although AtR and AtR(a) are different dynamically, they achieve comparable $\chi^2$ in Table.\ref{chi2}, implying that current data is not sufficient to distinguish both cases in our model. To further assess their difference and potential signatures to distinguish them, we examine their perturbation evolution. In the synchronous gauge, the scalar-field perturbation obeys
\be
\delta\phi_k''+2\mathcal{H}\delta\phi_k'+\left(k^2+a^2V_{,\phi\phi}\right)\delta\phi_k=-\frac{1}{2}\phi'h_k'.
\label{perturbation}
\ee
For the attractor branch $V_-=(\gamma\alpha^6/8)m_a^2f_a^2\tanh^6x$, with $x\equiv\phi/(\alpha f_a)$, the effective mass is
\be
V_{,\phi\phi}=(3/4)\gamma\alpha^4 m_a^2\tanh^4x\,\mathrm{sech}^2x\left(5-7\tanh^2x\right).
\label{effective_mass}
\ee
It becomes negative for $\tanh^2x>5/7$, allowing modes satisfying $k^2/a^2+V_{,\phi\phi}<0$ to undergo a transient tachyonic enhancement. Along the approximately constant $\lambda$ direction in Eq.~(\ref{deg}), the effective mass scales as $\alpha^{-2}$. This scaling, together with the sensitive dependence of the field trajectory on $\alpha$ and $\gamma$, enriches the perturbation
dynamics of the low-$\alpha$ branch. To characterize the resulting metric response, we plot the relative residual of the Weyl potential,
\be
\frac{\Delta\Phi_W}{\Phi_W},
\qquad
\Delta\Phi_W\equiv \Phi_{W,\mathrm{model}}-\Phi_{W,\mathrm{AxionEDE}},
\ee
in Fig.~\ref{weyl}. The reflection produces a pronounced transient feature around the EDE epoch and leaves a negative correction that persists toward later times. The double-injection, on the other hand, modifies the Weyl potential over a broader redshift interval and generates a residual with the opposite sign over part of the evolution. Unlike AtR(a) and the AxionEDE reference, the bestfit AtR experiences sharp reflection at the $\phi<0$ branch, receiving less tachyonic enhancement thus has reduced $\Phi_{\rm W}$ amplitude at high $k$ in Fig.\ref{weyl}. The difference in perturbation evolution can be encoded into small-scale CMB anisotropies, see Fig.\ref{cmb}, providing potential observational discriminant between the two mechanisms.

\textbf{\textit{Conclusions.}} ---
In this paper, we have shown that the EDE tail provides an observable window into the underlying scalar potential that is not captured by the conventional peak parameters $f_{\rm EDE}$ and $z_c$ alone. To illustrate this point, we constructed a multiscale potential in which an initially axion-like evolution transitions to an $\alpha$-attractor-like branch after the primary energy injection. Although both sides approach the same $\phi^6$ behavior near the minimum, their large-field structures generate qualitatively different EDE tails, including reflection, transient slow roll, runaway evolution, and asymmetric oscillatory dilution.

Confronting this model with Planck NPIPE high-$\ell$ CMB, Planck 2018 low-$\ell$ and lensing, DESI DR2 BAO, and the combined PantheonPlusSH0ES data, we identify a low-$\alpha$ profile-likelihood minimum in which the EDE tail is reshaped by an $\alpha$-attractor-like slope. Relative to conventional AxionEDE, the field undergoes a steeper reflection, thereby modifying both the post-peak dilution history and its perturbative signatures. The profile likelihood reaches an improvement of $\Delta\chi^2=-3.34$ over AxionEDE, with the gain arising predominantly from the high-$\ell$ CMB likelihood $\Delta\chi^2_{\text{high-}\ell}=-2.31$, which is notably better than the improvement of AxionEDE over $\Lambda$CDM $\Delta\chi^2_{\text{high-}\ell}=-0.34$. The best-fit AtR solution reaches $H_0\simeq72.2$km/s/Mpc, although the displacement of the marginalized $H_0$ posterior remains modest. 
% The absence of a comparable preference in the axion-to-axion control model further indicates that this improvement is associated with the structure of the post-peak potential rather than merely with the introduction of additional parameters.

A narrow region of the low-$\alpha$ branch also permits the field to enter a transient slow-roll phase, producing a secondary energy injection. An illustrative double injection solution achieves $\Delta\chi^2\simeq-2.9$ and $H_0\simeq72.3$km/s/Mpc, showing that such dynamics remain compatible with the present data. The reflection and slow-roll branches remains undistinguishable in the Planck CMB data studied in the main text, but can produce distinct Weyl-potential and CMB signatures at small scales, partly through the transient tachyonic evolution of scalar perturbations. These differences could be probed by future high-resolution CMB polarization measurements from experiments such as the Simons Observatory and CMB-S4. Our results therefore establish the EDE tail as an additional diagnostic of early-universe scalar dynamics. More broadly, they suggest that improving the phenomenology of EDE need not rely exclusively on changing the height or timing of its primary peak: observationally relevant freedom also resides in the EDE tail.

Our results should not be interpreted as evidence for the specific piecewise potential considered here. Rather, this potential serves as a simplified tool to realize and study different EDE tails. Actually, changing the potential shape is not the only way that will result in different EDE tails. Other possible mechanisms include modifying the scalar field dynamics by higher order operators \cite{Ye:2024kus}, modification of gravity \cite{Ye:2024zpk} and interaction with other species \cite{Wang:2024jug}, which we leave for future study.

% of a broader possibility suggested by the data: EDE dynamics may involve a characteristic change in the structure or scale of the scalar potential after the primary energy injection. Such a transition can reshape the EDE tail and produce an appreciable improvement in the fit, even when the conventional peak parameters remain similar. From this perspective, the relevant physical message is not the particular matching of an axion-like branch to an $\alpha$-attractor-like branch, but the observational sensitivity to changes in the potential structure across different field scales. The EDE tail therefore provides an additional diagnostic of early-universe scalar dynamics and a complementary direction for constructing viable EDE models beyond modifications of the peak amplitude and location alone.

\textbf{\textit{Acknowledgments.}} YSP is supported by NSFC, No.12475064, National Key Research and Development Program of China, No.2021YFC2203004, and the Fundamental
Research Funds for the Central Universities. GY acknowledges support from the Fundamental Research Funds for the Central Universities No.E6EQ0202X2. 

\clearpage
\begin{center}
	\textbf{\large Supplemental Material}
\end{center}

\appendix
\section{A. MCMC results}
\label{app:mcmc}

\begin{figure*}
	\includegraphics[width=2\columnwidth]{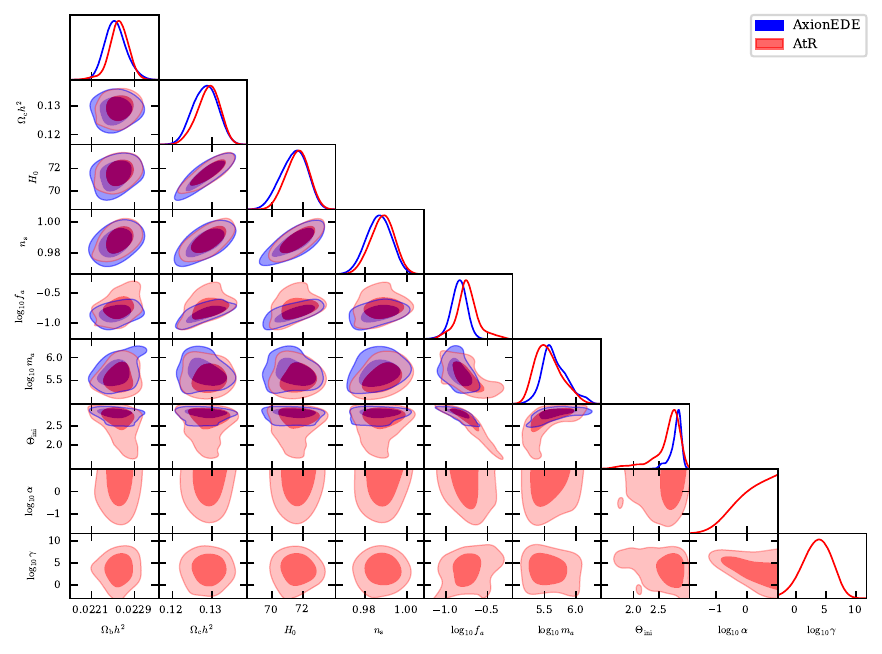}
	\caption{Marginalized posterior distributions for selected cosmological and scalar-field parameters in AxionEDE and AtR. The contours enclose the 68\% and 95\% credible regions. }
	\label{mc}
\end{figure*}

\section{B. Double-peak search method}
\label{app:double_peak}

Producing a secondary peak in the EDE tail requires finely tuned,
correlated choices of several potential parameters. The corresponding
region of parameter space can therefore be narrow and difficult
to sample adequately with a conventional MCMC analysis.
To investigate the observational constraints on this particular
shape of the EDE tail, we perform a dedicated search restricted
to histories with two distinct peaks of comparable amplitude.

\begin{figure}
	\includegraphics[width=1\columnwidth]{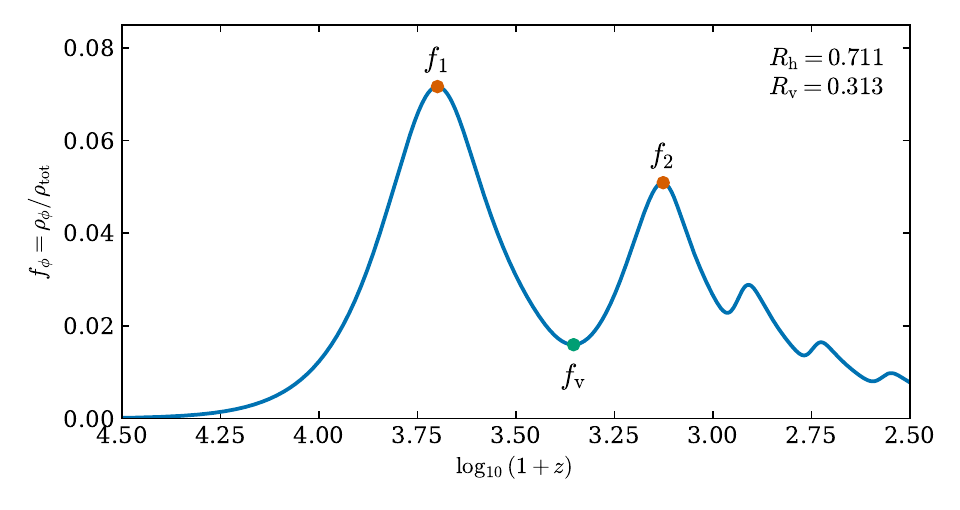}
	\caption{A representative double-peaked EDE evolution illustrating the selection criteria. Orange markers indicate the first two significant peaks, with amplitudes $f_1$ and $f_2$, and the green marker denotes the intervening minimum $f_{\rm v}$. This example satisfies $R_{\rm h}\equiv\min(f_1,f_2)/\max(f_1,f_2)\simeq0.711$ and $R_{\rm v}\equiv f_{\rm v}/\min(f_1,f_2)\simeq0.313$.}
	\label{dpmethod}
\end{figure}

We characterize the background evolution using the scalar-field
energy fraction,
\begin{equation}
	f_\phi(z)\equiv
	\frac{\rho_\phi(z)}{\rho_{\rm tot}(z)}.
\end{equation}
Significant local maxima are identified in chronological order,
with a minimum rise and fall of $10^{-4}$ in $f_\phi$ used
to distinguish peaks from small numerical fluctuations.
The first two qualifying maxima define the peak amplitudes
$f_1$ and $f_2$, while the minimum between them defines
the valley amplitude $f_{\rm v}$. We require
\begin{equation}
	\frac{\min(f_1,f_2)}{\max(f_1,f_2)}>0.7,
	\qquad
	\frac{f_{\rm v}}{\min(f_1,f_2)}<0.6.
\end{equation}
These conditions select peaks of comparable height separated
by a pronounced valley. They are applied to the first two
significant peaks, without replacing a rejected pair with
a later one, as illustrated in Fig.~\ref{dpmethod}.

The selection is implemented directly in the background
calculation: parameter points that fail either condition
are rejected before the CMB power spectra are computed.
We retain the same observational likelihoods as in the
unrestricted analysis and vary the cosmological and nuisance
parameters simultaneously. No additional restriction is imposed
on the peak redshifts or their separation. This procedure targets the double-peaked subset of the model and allows us to search for its best-fitting configurations.

% \section{C. CMB power spectra}
% \label{app:cmb_spectra}

% To illustrate the observable effects of the modified EDE tail, we compare the CMB temperature and polarization spectra of representative AtR configurations, labelled
% AtR and AtR(a), relative to a AxionEDE reference model. We define
% \begin{equation}
% 	D_\ell^{XY}\equiv
% 	\frac{\ell(\ell+1)}{2\pi}C_\ell^{XY},
% 	\qquad XY\in\{TT, EE\},
% \end{equation}
% and the absolute residual,
% \begin{equation}
% 	\Delta D_\ell^{XY}
% 	\equiv D_{\ell,{\rm model}}^{XY}
% 	-D_{\ell,{\rm ref}}^{XY}.
% \end{equation}

% Figure.~\ref{cmb} shows the relative residual $\Delta D_\ell^{XY}/D_{\ell,{\rm ref}}^{XY}$ of the lensed TT and EE spectra, together with the corresponding binned Planck NPIPE measurements.
% The data residuals are evaluated relative to the same
% reference spectra as the theoretical curves.
% The measurements are corrected using the adopted
% foreground and calibration parameters, as applicable,
% and their uncertainties are propagated through the binning
% using the covariance matrices. The statistical quality of each configuration is assessed
% using the full likelihood rather than the visual size
% of the plotted residuals.

\bibliographystyle{apsrev4-2}
\bibliography{references}

@article{Verde:2019ivm,
    author = "Verde, L. and Treu, T. and Riess, A. G.",
    title = "{Tensions between the Early and the Late Universe}",
    eprint = "1907.10625",
    archivePrefix = "arXiv",
    primaryClass = "astro-ph.CO",
    doi = "10.1038/s41550-019-0902-0",
    journal = "Nature Astron.",
    volume = "3",
    pages = "891",
    year = "2019"
}

@article{DiValentino:2020zio,
    author = "Di Valentino, Eleonora and others",
    title = "{Snowmass2021 - Letter of interest cosmology intertwined II: The hubble constant tension}",
    eprint = "2008.11284",
    archivePrefix = "arXiv",
    primaryClass = "astro-ph.CO",
    reportNumber = "FERMILAB-PUB-21-590-PPD",
    doi = "10.1016/j.astropartphys.2021.102605",
    journal = "Astropart. Phys.",
    volume = "131",
    pages = "102605",
    year = "2021"
}

@article{DiValentino:2021izs,
    author = "Di Valentino, Eleonora and Mena, Olga and Pan, Supriya and Visinelli, Luca and Yang, Weiqiang and Melchiorri, Alessandro and Mota, David F. and Riess, Adam G. and Silk, Joseph",
    title = "{In the realm of the Hubble tension{\textemdash}a review of solutions}",
    eprint = "2103.01183",
    archivePrefix = "arXiv",
    primaryClass = "astro-ph.CO",
    reportNumber = "IPPP/20/108",
    doi = "10.1088/1361-6382/ac086d",
    journal = "Class. Quant. Grav.",
    volume = "38",
    number = "15",
    pages = "153001",
    year = "2021"
}

@article{Perivolaropoulos:2021jda,
    author = "Perivolaropoulos, Leandros and Skara, Foteini",
    title = "{Challenges for {\ensuremath{\Lambda}}CDM: An update}",
    eprint = "2105.05208",
    archivePrefix = "arXiv",
    primaryClass = "astro-ph.CO",
    doi = "10.1016/j.newar.2022.101659",
    journal = "New Astron. Rev.",
    volume = "95",
    pages = "101659",
    year = "2022"
}

@article{Shah:2021onj,
    author = "Shah, Paul and Lemos, Pablo and Lahav, Ofer",
    title = "{A buyer{\textquoteright}s guide to the Hubble constant}",
    eprint = "2109.01161",
    archivePrefix = "arXiv",
    primaryClass = "astro-ph.CO",
    doi = "10.1007/s00159-021-00137-4",
    journal = "Astron. Astrophys. Rev.",
    volume = "29",
    number = "1",
    pages = "9",
    year = "2021"
}

@article{Abdalla:2022yfr,
    author = "Abdalla, Elcio and others",
    title = "{Cosmology intertwined: A review of the particle physics, astrophysics, and cosmology associated with the cosmological tensions and anomalies}",
    eprint = "2203.06142",
    archivePrefix = "arXiv",
    primaryClass = "astro-ph.CO",
    reportNumber = "FERMILAB-CONF-22-192-SCD",
    doi = "10.1016/j.jheap.2022.04.002",
    journal = "JHEAp",
    volume = "34",
    pages = "49--211",
    year = "2022"
}

@article{Hu:2023jqc,
    author = "Hu, Jian-Ping and Wang, Fa-Yin",
    title = "{Hubble Tension: The Evidence of New Physics}",
    eprint = "2302.05709",
    archivePrefix = "arXiv",
    primaryClass = "astro-ph.CO",
    doi = "10.3390/universe9020094",
    journal = "Universe",
    volume = "9",
    number = "2",
    pages = "94",
    year = "2023"
}

@article{Verde:2023lmm,
    author = {Verde, Licia and Sch{\"o}neberg, Nils and Gil-Mar{\'\i}n, H{\'e}ctor},
    title = "{A Tale of Many H0}",
    eprint = "2311.13305",
    archivePrefix = "arXiv",
    primaryClass = "astro-ph.CO",
    doi = "10.1146/annurev-astro-052622-033813",
    journal = "Ann. Rev. Astron. Astrophys.",
    volume = "62",
    number = "1",
    pages = "287--331",
    year = "2024"
}

@article{CosmoVerseNetwork:2025alb,
    author = "Di Valentino, Eleonora and others",
    collaboration = "CosmoVerse Network",
    title = "{The CosmoVerse White Paper: Addressing observational tensions in cosmology with systematics and fundamental physics}",
    eprint = "2504.01669",
    archivePrefix = "arXiv",
    primaryClass = "astro-ph.CO",
    doi = "10.1016/j.dark.2025.101965",
    journal = "Phys. Dark Univ.",
    volume = "49",
    pages = "101965",
    year = "2025"
}

@article{H0DN:2025lyy,
    author = "Casertano, Stefano and others",
    collaboration = "H0DN",
    title = "{The Local Distance Network: A community consensus report on the measurement of the Hubble constant at {\ensuremath{\sim}}1{\%} precision}",
    eprint = "2510.23823",
    archivePrefix = "arXiv",
    primaryClass = "astro-ph.CO",
    doi = "10.1051/0004-6361/202557993",
    journal = "Astron. Astrophys.",
    volume = "708",
    pages = "A166",
    year = "2026"
}

@article{SPT-3G:2025bzu,
    author = "Camphuis, E. and others",
    collaboration = "SPT-3G",
    title = "{SPT-3G D1: CMB temperature and polarization power spectra and cosmology from 2019 and 2020 observations of the SPT-3G main field}",
    eprint = "2506.20707",
    archivePrefix = "arXiv",
    primaryClass = "astro-ph.CO",
    reportNumber = "FERMILAB-PUB-25-0144-PPD",
    doi = "10.1103/7wt3-9v2y",
    journal = "Phys. Rev. D",
    volume = "113",
    number = "8",
    pages = "083504",
    year = "2026"
}

@article{Cuceu:2019for,
    author = "Cuceu, Andrei and Farr, James and Lemos, Pablo and Font-Ribera, Andreu",
    title = "{Baryon Acoustic Oscillations and the Hubble Constant: Past, Present and Future}",
    eprint = "1906.11628",
    archivePrefix = "arXiv",
    primaryClass = "astro-ph.CO",
    doi = "10.1088/1475-7516/2019/10/044",
    journal = "JCAP",
    volume = "10",
    pages = "044",
    year = "2019"
}

@article{Schoneberg:2019wmt,
    author = {Sch{\"o}neberg, Nils and Lesgourgues, Julien and Hooper, Deanna C.},
    title = "{The BAO+BBN take on the Hubble tension}",
    eprint = "1907.11594",
    archivePrefix = "arXiv",
    primaryClass = "astro-ph.CO",
    reportNumber = "TTK-19-30",
    doi = "10.1088/1475-7516/2019/10/029",
    journal = "JCAP",
    volume = "10",
    pages = "029",
    year = "2019"
}

@article{DESI:2025zgx,
    author = "Abdul Karim, M. and others",
    collaboration = "DESI",
    title = "{DESI DR2 results. II. Measurements of baryon acoustic oscillations and cosmological constraints}",
    eprint = "2503.14738",
    archivePrefix = "arXiv",
    primaryClass = "astro-ph.CO",
    reportNumber = "FERMILAB-PUB-25-0169-PPD",
    doi = "10.1103/tr6y-kpc6",
    journal = "Phys. Rev. D",
    volume = "112",
    number = "8",
    pages = "083515",
    year = "2025"
}

@article{Efstathiou:2020wxn,
    journal = {},
    author = "Efstathiou, G.",
    title = "{A Lockdown Perspective on the Hubble Tension (with comments from the SH0ES team)}",
    eprint = "2007.10716",
    archivePrefix = "arXiv",
    primaryClass = "astro-ph.CO",
    month = "7",
    year = "2020"
}

@article{Mortsell:2021tcx,
    author = "Mortsell, Edvard and Goobar, Ariel and Johansson, Joel and Dhawan, Suhail",
    title = "{The Hubble Tension Revisited: Additional Local Distance Ladder Uncertainties}",
    eprint = "2106.09400",
    archivePrefix = "arXiv",
    primaryClass = "astro-ph.CO",
    doi = "10.3847/1538-4357/ac7c19",
    journal = "Astrophys. J.",
    volume = "935",
    number = "1",
    pages = "58",
    year = "2022"
}

@article{Kenworthy:2022jdh,
    author = "Kenworthy, W. D'Arcy and Riess, Adam G. and Scolnic, Daniel and Yuan, Wenlong and Bernal, Jos{\'e} Luis and Brout, Dillon and Cassertano, Stefano and Jones, David O. and Macri, Lucas and Peterson, Erik R.",
    title = "{Measurements of the Hubble Constant with a Two-rung Distance Ladder: Two Out of Three Ain{\textquoteright}t Bad}",
    eprint = "2204.10866",
    archivePrefix = "arXiv",
    primaryClass = "astro-ph.CO",
    doi = "10.3847/1538-4357/ac80bd",
    journal = "Astrophys. J.",
    volume = "935",
    number = "2",
    pages = "83",
    year = "2022"
}

@article{Camarena:2022iae,
    author = "Camarena, David and Marra, Valerio and Sakr, Ziad and Clarkson, Chris",
    title = "{A void in the Hubble tension? The end of the line for the Hubble bubble}",
    eprint = "2205.05422",
    archivePrefix = "arXiv",
    primaryClass = "astro-ph.CO",
    doi = "10.1088/1361-6382/ac8635",
    journal = "Class. Quant. Grav.",
    volume = "39",
    number = "18",
    pages = "184001",
    year = "2022"
}

@article{Wojtak:2022bct,
    author = "Wojtak, Rados{\l}aw and Hjorth, Jens",
    title = "{Intrinsic tension in the supernova sector of the local Hubble constant measurement and its implications}",
    eprint = "2206.08160",
    archivePrefix = "arXiv",
    primaryClass = "astro-ph.CO",
    doi = "10.1093/mnras/stac1878",
    journal = "Mon. Not. Roy. Astron. Soc.",
    volume = "515",
    number = "2",
    pages = "2790--2799",
    year = "2022"
}

@article{Riess:2023bfx,
    author = "Riess, Adam G. and Anand, Gagandeep S. and Yuan, Wenlong and Casertano, Stefano and Dolphin, Andrew and Macri, Lucas M. and Breuval, Louise and Scolnic, Dan and Perrin, Marshall and Anderson, Richard I.",
    title = "{Crowded No More: The Accuracy of the Hubble Constant Tested with High-resolution Observations of Cepheids by JWST}",
    eprint = "2307.15806",
    archivePrefix = "arXiv",
    primaryClass = "astro-ph.CO",
    doi = "10.3847/2041-8213/acf769",
    journal = "Astrophys. J. Lett.",
    volume = "956",
    number = "1",
    pages = "L18",
    year = "2023"
}

@article{Giani:2023aor,
    author = "Giani, Leonardo and Howlett, Cullan and Said, Khaled and Davis, Tamara and Vagnozzi, Sunny",
    title = "{An effective description of Laniakea: impact on cosmology and the local determination of the Hubble constant}",
    eprint = "2311.00215",
    archivePrefix = "arXiv",
    primaryClass = "astro-ph.CO",
    doi = "10.1088/1475-7516/2024/01/071",
    journal = "JCAP",
    volume = "01",
    pages = "071",
    year = "2024"
}

@article{Wojtak:2024mgg,
    author = "Wojtak, Rados{\l}aw and Hjorth, Jens",
    title = "{Consistent extinction model for type Ia supernovae in Cepheid-based calibration galaxies and its impact on H0}",
    eprint = "2403.10388",
    archivePrefix = "arXiv",
    primaryClass = "astro-ph.CO",
    doi = "10.1093/mnras/stae1977",
    journal = "Mon. Not. Roy. Astron. Soc.",
    volume = "533",
    number = "2",
    pages = "2319--2334",
    year = "2024"
}

@article{Freedman:2024eph,
    author = "Freedman, Wendy L. and Madore, Barry F. and Hoyt, Taylor J. and Jang, In Sung and Lee, Abigail J. and Owens, Kayla A.",
    title = "{Status Report on the Chicago-Carnegie Hubble Program (CCHP): Measurement of the Hubble Constant Using the Hubble and James Webb Space Telescopes}",
    eprint = "2408.06153",
    archivePrefix = "arXiv",
    primaryClass = "astro-ph.CO",
    doi = "10.3847/1538-4357/adce78",
    journal = "Astrophys. J.",
    volume = "985",
    number = "2",
    pages = "203",
    year = "2025",
    note = "[Erratum: Astrophys.J. 993, 252 (2025)]"
}

@article{Perivolaropoulos:2024yxv,
    author = "Perivolaropoulos, Leandros",
    title = "{Hubble tension or distance ladder crisis?}",
    eprint = "2408.11031",
    archivePrefix = "arXiv",
    primaryClass = "astro-ph.CO",
    doi = "10.1103/PhysRevD.110.123518",
    journal = "Phys. Rev. D",
    volume = "110",
    number = "12",
    pages = "123518",
    year = "2024"
}

@article{Riess:2025chq,
    author = "Riess, Adam G. and others",
    title = "{The Perfect Host: JWST Cepheid Observations in a Background-free Type Ia Supernova Host Confirm No Bias in Hubble-constant Measurements}",
    eprint = "2509.01667",
    archivePrefix = "arXiv",
    primaryClass = "astro-ph.CO",
    doi = "10.3847/2041-8213/ae0ad6",
    journal = "Astrophys. J. Lett.",
    volume = "992",
    number = "2",
    pages = "L34",
    year = "2025"
}

@article{Hogas:2026urs,
    author = {H{\"o}g{\r{a}}s, Marcus and M{\"o}rtsell, Edvard},
    title = "{Physically motivated priors in the local distance ladder significantly reduce the Hubble tension}",
    eprint = "2601.22215",
    archivePrefix = "arXiv",
    primaryClass = "astro-ph.CO",
    doi = "10.1093/mnras/stag724",
    journal = "Mon. Not. Roy. Astron. Soc.",
    volume = "548",
    number = "3",
    pages = "stag724",
    year = "2026"
}

@article{Mortsell:2018mfj,
    author = {M{\"o}rtsell, Edvard and Dhawan, Suhail},
    title = "{Does the Hubble constant tension call for new physics?}",
    eprint = "1801.07260",
    archivePrefix = "arXiv",
    primaryClass = "astro-ph.CO",
    doi = "10.1088/1475-7516/2018/09/025",
    journal = "JCAP",
    volume = "09",
    pages = "025",
    year = "2018"
}

@article{Vagnozzi:2018jhn,
    author = "Vagnozzi, Sunny and Dhawan, Suhail and Gerbino, Martina and Freese, Katherine and Goobar, Ariel and Mena, Olga",
    title = "{Constraints on the sum of the neutrino masses in dynamical dark energy models with $w(z) \geq -1$ are tighter than those obtained in $\Lambda$CDM}",
    eprint = "1801.08553",
    archivePrefix = "arXiv",
    primaryClass = "astro-ph.CO",
    reportNumber = "LCTP-18-03,NORDITA-2018-004,IFIC/18-02, LCTP-18-03, NORDITA-2018-004, IFIC/18-02, IFIC-18-02",
    doi = "10.1103/PhysRevD.98.083501",
    journal = "Phys. Rev. D",
    volume = "98",
    number = "8",
    pages = "083501",
    year = "2018"
}

@article{Yang:2018euj,
    author = "Yang, Weiqiang and Pan, Supriya and Di Valentino, Eleonora and Nunes, Rafael C. and Vagnozzi, Sunny and Mota, David F.",
    title = "{Tale of stable interacting dark energy, observational signatures, and the $H_0$ tension}",
    eprint = "1805.08252",
    archivePrefix = "arXiv",
    primaryClass = "astro-ph.CO",
    doi = "10.1088/1475-7516/2018/09/019",
    journal = "JCAP",
    volume = "09",
    pages = "019",
    year = "2018"
}

@article{Guo:2018ans,
    author = "Guo, Rui-Yun and Zhang, Jing-Fei and Zhang, Xin",
    title = "{Can the $H_0$ tension be resolved in extensions to $\Lambda$CDM cosmology?}",
    eprint = "1809.02340",
    archivePrefix = "arXiv",
    primaryClass = "astro-ph.CO",
    doi = "10.1088/1475-7516/2019/02/054",
    journal = "JCAP",
    volume = "02",
    pages = "054",
    year = "2019"
}

@article{Kreisch:2019yzn,
    author = "Kreisch, Christina D. and Cyr-Racine, Francis-Yan and Dor{\'e}, Olivier",
    title = "{Neutrino puzzle: Anomalies, interactions, and cosmological tensions}",
    eprint = "1902.00534",
    archivePrefix = "arXiv",
    primaryClass = "astro-ph.CO",
    doi = "10.1103/PhysRevD.101.123505",
    journal = "Phys. Rev. D",
    volume = "101",
    number = "12",
    pages = "123505",
    year = "2020"
}

@article{Vagnozzi:2019ezj,
    author = "Vagnozzi, Sunny",
    title = "{New physics in light of the $H_0$ tension: An alternative view}",
    eprint = "1907.07569",
    archivePrefix = "arXiv",
    primaryClass = "astro-ph.CO",
    doi = "10.1103/PhysRevD.102.023518",
    journal = "Phys. Rev. D",
    volume = "102",
    number = "2",
    pages = "023518",
    year = "2020"
}

@article{Visinelli:2019qqu,
    author = "Visinelli, Luca and Vagnozzi, Sunny and Danielsson, Ulf",
    title = "{Revisiting a negative cosmological constant from low-redshift data}",
    eprint = "1907.07953",
    archivePrefix = "arXiv",
    primaryClass = "astro-ph.CO",
    doi = "10.3390/sym11081035",
    journal = "Symmetry",
    volume = "11",
    number = "8",
    pages = "1035",
    year = "2019"
}

@article{DiValentino:2019ffd,
    author = "Di Valentino, Eleonora and Melchiorri, Alessandro and Mena, Olga and Vagnozzi, Sunny",
    title = "{Interacting dark energy in the early 2020s: A promising solution to the $H_0$ and cosmic shear tensions}",
    eprint = "1908.04281",
    archivePrefix = "arXiv",
    primaryClass = "astro-ph.CO",
    doi = "10.1016/j.dark.2020.100666",
    journal = "Phys. Dark Univ.",
    volume = "30",
    pages = "100666",
    year = "2020"
}

@article{DiValentino:2019jae,
    author = "Di Valentino, Eleonora and Melchiorri, Alessandro and Mena, Olga and Vagnozzi, Sunny",
    title = "{Nonminimal dark sector physics and cosmological tensions}",
    eprint = "1910.09853",
    archivePrefix = "arXiv",
    primaryClass = "astro-ph.CO",
    doi = "10.1103/PhysRevD.101.063502",
    journal = "Phys. Rev. D",
    volume = "101",
    number = "6",
    pages = "063502",
    year = "2020"
}

@article{Krishnan:2020obg,
    author = "Krishnan, C. and Colg{\'a}in, Eoin {\'O}. and Ruchika and Sen, Anjan A. and Sheikh-Jabbari, M. M. and Yang, Tao",
    title = "{Is there an early Universe solution to Hubble tension?}",
    eprint = "2002.06044",
    archivePrefix = "arXiv",
    primaryClass = "astro-ph.CO",
    doi = "10.1103/PhysRevD.102.103525",
    journal = "Phys. Rev. D",
    volume = "102",
    number = "10",
    pages = "103525",
    year = "2020"
}

@article{Alestas:2020mvb,
    author = "Alestas, G. and Kazantzidis, L. and Perivolaropoulos, L.",
    title = "{$H_0$ tension, phantom dark energy, and cosmological parameter degeneracies}",
    eprint = "2004.08363",
    archivePrefix = "arXiv",
    primaryClass = "astro-ph.CO",
    doi = "10.1103/PhysRevD.101.123516",
    journal = "Phys. Rev. D",
    volume = "101",
    number = "12",
    pages = "123516",
    year = "2020"
}

@article{Jedamzik:2020krr,
    author = "Jedamzik, Karsten and Pogosian, Levon",
    title = "{Relieving the Hubble tension with primordial magnetic fields}",
    eprint = "2004.09487",
    archivePrefix = "arXiv",
    primaryClass = "astro-ph.CO",
    doi = "10.1103/PhysRevLett.125.181302",
    journal = "Phys. Rev. Lett.",
    volume = "125",
    number = "18",
    pages = "181302",
    year = "2020"
}

@article{Sekiguchi:2020teg,
    author = "Sekiguchi, Toyokazu and Takahashi, Tomo",
    title = "{Early recombination as a solution to the $H_0$ tension}",
    eprint = "2007.03381",
    archivePrefix = "arXiv",
    primaryClass = "astro-ph.CO",
    reportNumber = "KEK-TH-2238",
    doi = "10.1103/PhysRevD.103.083507",
    journal = "Phys. Rev. D",
    volume = "103",
    number = "8",
    pages = "083507",
    year = "2021"
}

@article{RoyChoudhury:2020dmd,
    author = "Roy Choudhury, Shouvik and Hannestad, Steen and Tram, Thomas",
    title = "{Updated constraints on massive neutrino self-interactions from cosmology in light of the $H_0$ tension}",
    eprint = "2012.07519",
    archivePrefix = "arXiv",
    primaryClass = "astro-ph.CO",
    doi = "10.1088/1475-7516/2021/03/084",
    journal = "JCAP",
    volume = "03",
    pages = "084",
    year = "2021"
}

@article{Brinckmann:2020bcn,
    author = "Brinckmann, Thejs and Chang, Jae Hyeok and LoVerde, Marilena",
    title = "{Self-interacting neutrinos, the Hubble parameter tension, and the cosmic microwave background}",
    eprint = "2012.11830",
    archivePrefix = "arXiv",
    primaryClass = "astro-ph.CO",
    reportNumber = "YITP-SB-2020-40",
    doi = "10.1103/PhysRevD.104.063523",
    journal = "Phys. Rev. D",
    volume = "104",
    number = "6",
    pages = "063523",
    year = "2021"
}

@article{Gao:2021xnk,
    author = "Gao, Li-Yang and Zhao, Ze-Wei and Xue, She-Sheng and Zhang, Xin",
    title = "{Relieving the H 0 tension with a new interacting dark energy model}",
    eprint = "2101.10714",
    archivePrefix = "arXiv",
    primaryClass = "astro-ph.CO",
    doi = "10.1088/1475-7516/2021/07/005",
    journal = "JCAP",
    volume = "07",
    pages = "005",
    year = "2021"
}

@article{Marra:2021fvf,
    author = "Marra, Valerio and Perivolaropoulos, Leandros",
    title = "{Rapid transition of Geff at zt{\ensuremath{\simeq}}0.01 as a possible solution of the Hubble and growth tensions}",
    eprint = "2102.06012",
    archivePrefix = "arXiv",
    primaryClass = "astro-ph.CO",
    doi = "10.1103/PhysRevD.104.L021303",
    journal = "Phys. Rev. D",
    volume = "104",
    number = "2",
    pages = "L021303",
    year = "2021"
}

@article{Dainotti:2021pqg,
    author = "Dainotti, Maria Giovanna and De Simone, Biagio and Schiavone, Tiziano and Montani, Giovanni and Rinaldi, Enrico and Lambiase, Gaetano",
    title = "{On the Hubble constant tension in the SNe Ia Pantheon sample}",
    eprint = "2103.02117",
    archivePrefix = "arXiv",
    primaryClass = "astro-ph.CO",
    reportNumber = "AAS28095R4, RIKEN-iTHEMS-Report-21",
    doi = "10.3847/1538-4357/abeb73",
    journal = "Astrophys. J.",
    volume = "912",
    number = "2",
    pages = "150",
    year = "2021"
}

@article{Krishnan:2021dyb,
    author = "Krishnan, Chethan and Mohayaee, Roya and Colg{\'a}in, Eoin {\'O}. and Sheikh-Jabbari, M. M. and Yin, Lu",
    title = "{Does Hubble tension signal a breakdown in FLRW cosmology?}",
    eprint = "2105.09790",
    archivePrefix = "arXiv",
    primaryClass = "astro-ph.CO",
    doi = "10.1088/1361-6382/ac1a81",
    journal = "Class. Quant. Grav.",
    volume = "38",
    number = "18",
    pages = "184001",
    year = "2021"
}

@article{Cyr-Racine:2021oal,
    author = "Cyr-Racine, Francis-Yan and Ge, Fei and Knox, Lloyd",
    title = "{Symmetry of Cosmological Observables, a Mirror World Dark Sector, and the Hubble Constant}",
    eprint = "2107.13000",
    archivePrefix = "arXiv",
    primaryClass = "astro-ph.CO",
    doi = "10.1103/PhysRevLett.128.201301",
    journal = "Phys. Rev. Lett.",
    volume = "128",
    number = "20",
    pages = "201301",
    year = "2022"
}

@article{Anchordoqui:2021gji,
    author = "Anchordoqui, Luis A. and Di Valentino, Eleonora and Pan, Supriya and Yang, Weiqiang",
    title = "{Dissecting the H0 and S8 tensions with Planck + BAO + supernova type Ia in multi-parameter cosmologies}",
    eprint = "2107.13932",
    archivePrefix = "arXiv",
    primaryClass = "astro-ph.CO",
    reportNumber = "IPPP/21/17",
    doi = "10.1016/j.jheap.2021.08.001",
    journal = "JHEAp",
    volume = "32",
    pages = "28--64",
    year = "2021"
}

@article{Akarsu:2021fol,
    author = {Akarsu, {\"O}zg{\"u}r and Kumar, Suresh and {\"O}z{\"u}lker, Emre and Vazquez, J. Alberto},
    title = "{Relaxing cosmological tensions with a sign switching cosmological constant}",
    eprint = "2108.09239",
    archivePrefix = "arXiv",
    primaryClass = "astro-ph.CO",
    doi = "10.1103/PhysRevD.104.123512",
    journal = "Phys. Rev. D",
    volume = "104",
    number = "12",
    pages = "123512",
    year = "2021"
}

@article{Ren:2022aeo,
    author = "Ren, Xin and Yan, Sheng-Feng and Zhao, Yaqi and Cai, Yi-Fu and Saridakis, Emmanuel N.",
    title = "{Gaussian processes and effective field theory of $f(T)$ gravity under the $H_0$ tension}",
    eprint = "2203.01926",
    archivePrefix = "arXiv",
    primaryClass = "astro-ph.CO",
    doi = "10.3847/1538-4357/ac6ba5",
    journal = "Astrophys. J.",
    volume = "932",
    number = "2",
    pages = "131",
    year = "2022"
}

@article{Nojiri:2022ski,
    author = "Nojiri, S. and Odintsov, S. D. and Oikonomou, V. K.",
    title = "{Integral F(R) gravity and saddle point condition as a remedy for the H0-tension}",
    eprint = "2205.11681",
    archivePrefix = "arXiv",
    primaryClass = "gr-qc",
    doi = "10.1016/j.nuclphysb.2022.115850",
    journal = "Nucl. Phys. B",
    volume = "980",
    pages = "115850",
    year = "2022"
}

@article{Schoneberg:2022grr,
    author = {Sch{\"o}neberg, Nils and Franco Abell{\'a}n, Guillermo},
    title = "{A step in the right direction? Analyzing the Wess Zumino Dark Radiation solution to the Hubble tension}",
    eprint = "2206.11276",
    archivePrefix = "arXiv",
    primaryClass = "astro-ph.CO",
    doi = "10.1088/1475-7516/2022/12/001",
    journal = "JCAP",
    volume = "12",
    pages = "001",
    year = "2022"
}

@article{Banerjee:2022ynv,
    author = "Banerjee, Shreya and Petronikolou, Maria and Saridakis, Emmanuel N.",
    title = "{Alleviating the H0 tension with new gravitational scalar tensor theories}",
    eprint = "2209.02426",
    archivePrefix = "arXiv",
    primaryClass = "gr-qc",
    doi = "10.1103/PhysRevD.108.024012",
    journal = "Phys. Rev. D",
    volume = "108",
    number = "2",
    pages = "024012",
    year = "2023"
}

@article{deSa:2022hsh,
    author = "de S{\'a}, Ramon and Benetti, Micol and Graef, Leila Lobato",
    title = "{An empirical investigation into cosmological tensions}",
    eprint = "2209.11476",
    archivePrefix = "arXiv",
    primaryClass = "astro-ph.CO",
    doi = "10.1140/epjp/s13360-022-03343-w",
    journal = "Eur. Phys. J. Plus",
    volume = "137",
    number = "10",
    pages = "1129",
    year = "2022"
}

@article{Akarsu:2022typ,
    author = {Akarsu, Ozgur and Kumar, Suresh and {\"O}z{\"u}lker, Emre and Vazquez, J. Alberto and Yadav, Anita},
    title = "{Relaxing cosmological tensions with a sign switching cosmological constant: Improved results with Planck, BAO, and Pantheon data}",
    eprint = "2211.05742",
    archivePrefix = "arXiv",
    primaryClass = "astro-ph.CO",
    doi = "10.1103/PhysRevD.108.023513",
    journal = "Phys. Rev. D",
    volume = "108",
    number = "2",
    pages = "023513",
    year = "2023"
}

@article{Lee:2022gzh,
    author = {Lee, Nanoom and Ali-Ha{\"\i}moud, Yacine and Sch{\"o}neberg, Nils and Poulin, Vivian},
    title = "{What It Takes to Solve the Hubble Tension through Modifications of Cosmological Recombination}",
    eprint = "2212.04494",
    archivePrefix = "arXiv",
    primaryClass = "astro-ph.CO",
    doi = "10.1103/PhysRevLett.130.161003",
    journal = "Phys. Rev. Lett.",
    volume = "130",
    number = "16",
    pages = "161003",
    year = "2023"
}

@article{Khodadi:2023ezj,
    author = "Khodadi, Mohsen and Schreck, Marco",
    title = "{Hubble tension as a guide for refining the early Universe: Cosmologies with explicit local Lorentz and diffeomorphism violation}",
    eprint = "2301.03883",
    archivePrefix = "arXiv",
    primaryClass = "gr-qc",
    doi = "10.1016/j.dark.2023.101170",
    journal = "Phys. Dark Univ.",
    volume = "39",
    pages = "101170",
    year = "2023"
}

@article{Bernui:2023byc,
    author = "Bernui, Armando and Di Valentino, Eleonora and Giar{\`e}, William and Kumar, Suresh and Nunes, Rafael C.",
    title = "{Exploring the H0 tension and the evidence for dark sector interactions from 2D BAO measurements}",
    eprint = "2301.06097",
    archivePrefix = "arXiv",
    primaryClass = "astro-ph.CO",
    doi = "10.1103/PhysRevD.107.103531",
    journal = "Phys. Rev. D",
    volume = "107",
    number = "10",
    pages = "103531",
    year = "2023"
}

@article{Ben-Dayan:2023rgt,
    author = "Ben-Dayan, Ido and Kumar, Utkarsh",
    title = "{Emergent Unparticles Dark Energy can restore cosmological concordance}",
    eprint = "2302.00067",
    archivePrefix = "arXiv",
    primaryClass = "astro-ph.CO",
    doi = "10.1088/1475-7516/2023/12/047",
    journal = "JCAP",
    volume = "12",
    pages = "047",
    year = "2023"
}

@article{Gomez-Valent:2023hov,
    author = "G{\'o}mez-Valent, Adri{\`a} and Mavromatos, Nick E. and Sol{\`a} Peracaula, Joan",
    title = "{Stringy running vacuum model and current tensions in cosmology}",
    eprint = "2305.15774",
    archivePrefix = "arXiv",
    primaryClass = "gr-qc",
    doi = "10.1088/1361-6382/ad0fb8",
    journal = "Class. Quant. Grav.",
    volume = "41",
    number = "1",
    pages = "015026",
    year = "2024"
}

@article{Ruchika:2023ugh,
    author = "Ruchika and Rathore, Himansh and Roy Choudhury, Shouvik and Rentala, Vikram",
    title = "{A gravitational constant transition within cepheids as supernovae calibrators can solve the Hubble tension}",
    eprint = "2306.05450",
    archivePrefix = "arXiv",
    primaryClass = "astro-ph.CO",
    doi = "10.1088/1475-7516/2024/06/056",
    journal = "JCAP",
    volume = "06",
    pages = "056",
    year = "2024"
}

@article{Adil:2023exv,
    author = {Adil, Shahnawaz A. and Akarsu, {\"O}zg{\"u}r and Di Valentino, Eleonora and Nunes, Rafael C. and {\"O}z{\"u}lker, Emre and Sen, Anjan A. and Specogna, Enrico},
    title = "{Omnipotent dark energy: A phenomenological answer to the Hubble tension}",
    eprint = "2306.08046",
    archivePrefix = "arXiv",
    primaryClass = "astro-ph.CO",
    doi = "10.1103/PhysRevD.109.023527",
    journal = "Phys. Rev. D",
    volume = "109",
    number = "2",
    pages = "023527",
    year = "2024"
}

@article{Frion:2023xwq,
    journal = {},
    author = "Frion, Emmanuel and Camarena, David and Giani, Leonardo and Miranda, Tays and Bertacca, Daniele and Marra, Valerio and Piattella, Oliver F.",
    title = "{Bayesian analysis of a Unified Dark Matter model with transition: can it alleviate the $H_{0}$ tension?}",
    eprint = "2307.06320",
    archivePrefix = "arXiv",
    primaryClass = "astro-ph.CO",
    doi = "10.21105/astro.2307.06320",
    month = "7",
    year = "2023"
}

@article{Gomez-Valent:2023uof,
    author = "G{\'o}mez-Valent, Adri{\`a} and Favale, Arianna and Migliaccio, Marina and Sen, Anjan A.",
    title = "{Late-time phenomenology required to solve the H0 tension in view of the cosmic ladders and the anisotropic and angular BAO datasets}",
    eprint = "2309.07795",
    archivePrefix = "arXiv",
    primaryClass = "astro-ph.CO",
    doi = "10.1103/PhysRevD.109.023525",
    journal = "Phys. Rev. D",
    volume = "109",
    number = "2",
    pages = "023525",
    year = "2024"
}

@article{Akarsu:2024qiq,
    author = {Akarsu, {\"O}zg{\"u}r and Colg{\'a}in, Eoin {\'O}. and Sen, Anjan A. and Sheikh-Jabbari, M. M.},
    title = "{{\ensuremath{\Lambda}}CDM Tensions: Localising Missing Physics through Consistency Checks}",
    eprint = "2402.04767",
    archivePrefix = "arXiv",
    primaryClass = "astro-ph.CO",
    doi = "10.3390/universe10080305",
    journal = "Universe",
    volume = "10",
    number = "8",
    pages = "305",
    year = "2024"
}

@article{Giare:2024ytc,
    author = "Giar{\`e}, William and Zhai, Yuejia and Pan, Supriya and Di Valentino, Eleonora and Nunes, Rafael C. and van de Bruck, Carsten",
    title = "{Tightening the reins on nonminimal dark sector physics: Interacting dark energy with dynamical and nondynamical equation of state}",
    eprint = "2404.02110",
    archivePrefix = "arXiv",
    primaryClass = "astro-ph.CO",
    doi = "10.1103/PhysRevD.110.063527",
    journal = "Phys. Rev. D",
    volume = "110",
    number = "6",
    pages = "063527",
    year = "2024"
}

@article{Lynch:2024gmp,
    author = "Lynch, Gabriel P. and Knox, Lloyd and Chluba, Jens",
    title = "{Reconstructing the recombination history by combining early and late cosmological probes}",
    eprint = "2404.05715",
    archivePrefix = "arXiv",
    primaryClass = "astro-ph.CO",
    doi = "10.1103/PhysRevD.110.063518",
    journal = "Phys. Rev. D",
    volume = "110",
    number = "6",
    pages = "063518",
    year = "2024"
}

@article{Giare:2024smz,
    author = "Giar{\`e}, William and Sabogal, Miguel A. and Nunes, Rafael C. and Di Valentino, Eleonora",
    title = "{Interacting Dark Energy after DESI Baryon Acoustic Oscillation Measurements}",
    eprint = "2404.15232",
    archivePrefix = "arXiv",
    primaryClass = "astro-ph.CO",
    doi = "10.1103/PhysRevLett.133.251003",
    journal = "Phys. Rev. Lett.",
    volume = "133",
    number = "25",
    pages = "251003",
    year = "2024"
}

@article{Lynch:2024hzh,
    author = "Lynch, Gabriel P. and Knox, Lloyd and Chluba, Jens",
    title = "{DESI observations and the Hubble tension in light of modified recombination}",
    eprint = "2406.10202",
    archivePrefix = "arXiv",
    primaryClass = "astro-ph.CO",
    doi = "10.1103/PhysRevD.110.083538",
    journal = "Phys. Rev. D",
    volume = "110",
    number = "8",
    pages = "083538",
    year = "2024"
}

@article{Toda:2024ncp,
    author = {Toda, Yo and Giar{\`e}, William and {\"O}z{\"u}lker, Emre and Di Valentino, Eleonora and Vagnozzi, Sunny},
    title = "{Combining pre- and post-recombination new physics to address cosmological tensions: Case study with varying electron mass and sign-switching cosmological constant}",
    eprint = "2407.01173",
    archivePrefix = "arXiv",
    primaryClass = "astro-ph.CO",
    doi = "10.1016/j.dark.2024.101676",
    journal = "Phys. Dark Univ.",
    volume = "46",
    pages = "101676",
    year = "2024"
}

@article{Nozari:2024wir,
    author = "Nozari, Kourosh and Saghafi, Sara and Hajebrahimi, Milad",
    title = "{Extended uncertainty principle: A deeper insight into the Hubble tension?}",
    eprint = "2407.01961",
    archivePrefix = "arXiv",
    primaryClass = "gr-qc",
    doi = "10.1016/j.dark.2024.101571",
    journal = "Phys. Dark Univ.",
    volume = "46",
    pages = "101571",
    year = "2024"
}

@article{Escamilla:2024xmz,
    author = "Escamilla, Luis A. and Fiorucci, Donatella and Montani, Giovanni and Di Valentino, Eleonora",
    title = "{Exploring the Hubble tension with a late time Modified Gravity scenario}",
    eprint = "2408.04354",
    archivePrefix = "arXiv",
    primaryClass = "astro-ph.CO",
    doi = "10.1016/j.dark.2024.101652",
    journal = "Phys. Dark Univ.",
    volume = "46",
    pages = "101652",
    year = "2024"
}

@article{RoyChoudhury:2024wri,
    author = "Roy Choudhury, Shouvik and Okumura, Teppei",
    title = "{Updated Cosmological Constraints in Extended Parameter Space with Planck PR4, DESI Baryon Acoustic Oscillations, and Supernovae: Dynamical Dark Energy, Neutrino Masses, Lensing Anomaly, and the Hubble Tension}",
    eprint = "2409.13022",
    archivePrefix = "arXiv",
    primaryClass = "astro-ph.CO",
    doi = "10.3847/2041-8213/ad8c26",
    journal = "Astrophys. J. Lett.",
    volume = "976",
    number = "1",
    pages = "L11",
    year = "2024"
}

@article{Mirpoorian:2024fka,
    author = "Mirpoorian, Seyed Hamidreza and Jedamzik, Karsten and Pogosian, Levon",
    title = "{Modified recombination and the Hubble tension}",
    eprint = "2411.16678",
    archivePrefix = "arXiv",
    primaryClass = "astro-ph.CO",
    doi = "10.1103/PhysRevD.111.083519",
    journal = "Phys. Rev. D",
    volume = "111",
    number = "8",
    pages = "083519",
    year = "2025"
}

@article{Li:2025owk,
    author = "Li, Tian-Nuo and Du, Guo-Hong and Li, Yun-He and Wu, Peng-Ju and Jin, Shang-Jie and Zhang, Jing-Fei and Zhang, Xin",
    title = "{Probing the sign-changeable interaction between dark energy and dark matter with DESI baryon acoustic oscillations and DES supernovae data}",
    eprint = "2501.07361",
    archivePrefix = "arXiv",
    primaryClass = "astro-ph.CO",
    doi = "10.1007/s11433-025-2771-5",
    journal = "Sci. China Phys. Mech. Astron.",
    volume = "69",
    number = "1",
    pages = "210413",
    year = "2026"
}

@article{Lee:2025yah,
    author = {Lee, Nanoom and Braglia, Matteo and Ali-Ha{\"\i}moud, Yacine},
    title = "{What it takes to solve the Hubble tension through scale-dependent modifications of the primordial power spectrum}",
    eprint = "2504.07966",
    archivePrefix = "arXiv",
    primaryClass = "astro-ph.CO",
    doi = "10.1103/9q3f-5zrd",
    journal = "Phys. Rev. D",
    volume = "112",
    number = "8",
    pages = "083506",
    year = "2025"
}

@article{Teixeira:2025czm,
    author = "Teixeira, Elsa M. and Giar{\`e}, William and Hogg, Natalie B. and Montandon, Thomas and Poudou, Ad{\`e}le and Poulin, Vivian",
    title = "{Implications of distance duality violation for the H0 tension and evolving dark energy}",
    eprint = "2504.10464",
    archivePrefix = "arXiv",
    primaryClass = "astro-ph.CO",
    doi = "10.1103/zzmp-rxrh",
    journal = "Phys. Rev. D",
    volume = "112",
    number = "2",
    pages = "023515",
    year = "2025"
}

@article{Wang:2025dzn,
    author = "Wang, Yi-Ying and Lei, Lei and Tang, Shao-Peng and Fan, Yi-Zhong",
    title = "{Lensing amplitude anomaly and varying electron mass alleviate the Hubble and S $_{8}$ tensions}",
    eprint = "2508.19081",
    archivePrefix = "arXiv",
    primaryClass = "astro-ph.CO",
    doi = "10.1088/1475-7516/2026/01/009",
    journal = "JCAP",
    volume = "01",
    pages = "009",
    year = "2026"
}

@article{Zhang:2025dwu,
    journal = {},
    author = "Zhang, Yi-Min and Li, Tian-Nuo and Du, Guo-Hong and Zhou, Sheng-Han and Gao, Li-Yang and Zhang, Jing-Fei and Zhang, Xin",
    title = "{Alleviating the $H_0$ tension through the interacting dark energy model from quantum gravitational field theory in light of DESI DR2}",
    eprint = "2510.12627",
    archivePrefix = "arXiv",
    primaryClass = "astro-ph.CO",
    month = "10",
    year = "2025"
}

@article{Kumar:2025obb,
    author = "Kumar, Suresh",
    title = "{{\ensuremath{\Omega}}1{\ensuremath{\Omega}}2{\textendash}{\ensuremath{\Lambda}}CDM: A promising phenomenological extension of the standard model of cosmology}",
    eprint = "2512.19000",
    archivePrefix = "arXiv",
    primaryClass = "astro-ph.CO",
    doi = "10.1016/j.dark.2026.102248",
    journal = "Phys. Dark Univ.",
    volume = "52",
    pages = "102248",
    year = "2026"
}

@article{Hogas:2025mii,
    author = {H{\"o}g{\r{a}}s, Marcus and M{\"o}rtsell, Edvard and Desmond, Harry and Riess, Adam},
    title = "{Constraining fifth forces using the local distance ladder: Implications for the Hubble tension}",
    eprint = "2512.14814",
    archivePrefix = "arXiv",
    primaryClass = "astro-ph.CO",
    doi = "10.1103/j4yp-9m6h",
    journal = "Phys. Rev. D",
    volume = "114",
    number = "2",
    pages = "023519",
    year = "2026"
}

@article{Pantos:2026rpe,
    author = "Pantos, Ioannis and Perivolaropoulos, Leandros",
    title = "{On the origin of the BAOtr-DESI tension}",
    eprint = "2604.11106",
    archivePrefix = "arXiv",
    primaryClass = "astro-ph.CO",
    doi = "10.1016/j.dark.2026.102347",
    journal = "Phys. Dark Univ.",
    volume = "52",
    pages = "102347",
    year = "2026"
}

@article{Bernal:2016gxb,
    author = "Bernal, Jose Luis and Verde, Licia and Riess, Adam G.",
    title = "{The trouble with $H_0$}",
    eprint = "1607.05617",
    archivePrefix = "arXiv",
    primaryClass = "astro-ph.CO",
    doi = "10.1088/1475-7516/2016/10/019",
    journal = "JCAP",
    volume = "10",
    pages = "019",
    year = "2016"
}

@article{Lemos:2018smw,
    author = "Lemos, Pablo and Lee, Elizabeth and Efstathiou, George and Gratton, Steven",
    title = "{Model independent $H(z)$ reconstruction using the cosmic inverse distance ladder}",
    eprint = "1806.06781",
    archivePrefix = "arXiv",
    primaryClass = "astro-ph.CO",
    doi = "10.1093/mnras/sty3082",
    journal = "Mon. Not. Roy. Astron. Soc.",
    volume = "483",
    number = "4",
    pages = "4803--4810",
    year = "2019"
}

@article{Aylor:2018drw,
    author = "Aylor, Kevin and Joy, MacKenzie and Knox, Lloyd and Millea, Marius and Raghunathan, Srinivasan and Wu, W. L. Kimmy",
    title = "{Sounds Discordant: Classical Distance Ladder {\&} $\Lambda$CDM -based Determinations of the Cosmological Sound Horizon}",
    eprint = "1811.00537",
    archivePrefix = "arXiv",
    primaryClass = "astro-ph.CO",
    doi = "10.3847/1538-4357/ab0898",
    journal = "Astrophys. J.",
    volume = "874",
    number = "1",
    pages = "4",
    year = "2019"
}

@article{Knox:2019rjx,
    author = "Knox, Lloyd and Millea, Marius",
    title = "{Hubble constant hunter{\textquoteright}s guide}",
    eprint = "1908.03663",
    archivePrefix = "arXiv",
    primaryClass = "astro-ph.CO",
    doi = "10.1103/PhysRevD.101.043533",
    journal = "Phys. Rev. D",
    volume = "101",
    number = "4",
    pages = "043533",
    year = "2020"
}

@article{Arendse:2019hev,
    author = "Arendse, Nikki and others",
    title = "{Cosmic dissonance: are new physics or systematics behind a short sound horizon?}",
    eprint = "1909.07986",
    archivePrefix = "arXiv",
    primaryClass = "astro-ph.CO",
    doi = "10.1051/0004-6361/201936720",
    journal = "Astron. Astrophys.",
    volume = "639",
    pages = "A57",
    year = "2020"
}

@article{Cai:2021weh,
    author = "Cai, Rong-Gen and Guo, Zong-Kuan and Wang, Shao-Jiang and Yu, Wang-Wei and Zhou, Yong",
    title = "{No-go guide for the Hubble tension: Late-time solutions}",
    eprint = "2107.13286",
    archivePrefix = "arXiv",
    primaryClass = "astro-ph.CO",
    doi = "10.1103/PhysRevD.105.L021301",
    journal = "Phys. Rev. D",
    volume = "105",
    number = "2",
    pages = "L021301",
    year = "2022"
}

@article{Zhou:2026jmn,
    journal = {},
    author = "Zhou, Quan and Zheng, Sibo",
    title = "{Revisiting the Hubble tension with an inverse power-law early dark energy}",
    eprint = "2608.26507",
    archivePrefix = "arXiv",
    primaryClass = "astro-ph.CO",
    month = "8",
    year = "2026"
}
\end{document}